\documentclass[a4paper,11pt]{article}

\pdfoutput=1 

\usepackage{jheppub} 

\usepackage{tikz}
\usetikzlibrary{patterns,arrows,decorations.pathmorphing,decorations.markings}

\usepackage{slashed,adjustbox,bm,mathrsfs,bbm}
\usepackage{subcaption}
\usepackage{cleveref}
\usepackage[T1]{fontenc} 
\usepackage[utf8]{inputenc}
\usepackage{mathtools}

\newcommand{\gA}{\mathscr{A}} 

\newcommand{\newD}{\mathscr{D}}
\newcommand{\nnewD}{ \widetilde{\mathscr{D}}}
\newcommand{\phiF}{Z}

\def\rd{{\rm d}}
\def\tr{{\rm Tr}\,}

\def\hyp{\mathsf{y}}
\def\nn{ \nonumber \\ }
\def\coef#1#2#3{ {\vphantom{A^2}}^{#1}\!{{C}}^{#3}_{#2} }
\def\dcoef#1#2#3{ {\vphantom{A^2}}^{#1}\!{\dot{{C}}}^{#3}_{#2} }
\def\hcoef#1#2#3{ {\vphantom{A^2}}^{#1}\!{{\widehat C}}^{#3}_{#2} }

\def\op#1#2#3{ {\vphantom{A^2}}^{#1}\!{Q}^{#3}_{#2} }
\def\opr#1#2#3{ {\vphantom{A^2}}^{#1}\!{R}^{#3}_{#2} }

\def\f#1#2#3{ f^{#1}_{\ #2#3}}

\makeatletter
\newsavebox{\@brx}
\newcommand{\llangle}[1][]{\savebox{\@brx}{\(\m@th{#1\langle}\)}%
  \mathopen{\copy\@brx\kern-0.5\wd\@brx\usebox{\@brx}}}
\newcommand{\rrangle}[1][]{\savebox{\@brx}{\(\m@th{#1\rangle}\)}%
  \mathclose{\copy\@brx\kern-0.5\wd\@brx\usebox{\@brx}}}
\makeatother

\subheader{\normalsize
	\vspace*{-3.7em}
	\begin{flushright}
		CALT-TH-2022-041
	\end{flushright}
}

\title{\boldmath Renormalization of the Standard Model Effective Field Theory from Geometry}

\author[a]{Andreas Helset,}

\affiliation[a]{Walter Burke Institute for Theoretical Physics,
California Institute of Technology,\\ Pasadena, CA 91125, USA}
\emailAdd{ahelset@caltech.edu}

\author[b]{Elizabeth E.~Jenkins,}

\affiliation[b]{Physics Department 0319,
University of California San Diego,\\ 9500 Gilman Drive, La Jolla, CA 92093-0319, USA}
\emailAdd{ejenkins@ucsd.edu}

\author[b]{and Aneesh V.~Manohar}

\emailAdd{amanohar@ucsd.edu}

\abstract{$S$-matrix elements are invariant under field redefinitions of the Lagrangian. They are determined by geometric quantities such as the curvature of the field-space manifold of scalar and gauge fields.  We present a formalism where scalar and gauge fields are treated together, with a metric on the combined space of both types of fields. Scalar and gauge scattering amplitudes are given by the Riemann curvature $R_{ijkl}$ of this combined space, with indices $i,j,k,l$ chosen to be scalar or gauge indices depending on the type of external particle. One-loop divergences can also be computed in terms of geometric invariants of the combined space, which greatly simplifies the computation of renormalization group equations. We apply our formalism to the Standard Model Effective Field Theory (SMEFT), and compute the renormalization group equations for even-parity bosonic operators to mass dimension eight.
}

\allowdisplaybreaks

\begin{document} 
\maketitle
\flushbottom

\section{Introduction}\label{sec:Introduction}

An important property of the $S$-matrix is its invariance under field redefinitions~\cite{Chisholm:1961tha,Kamefuchi:1961sb,Politzer:1980me,Arzt:1993gz}. The Lagrangian and correlation functions (Green's functions) change under field redefinitions; however, the scattering amplitudes and physical observables remain unchanged.\footnote{Scattering amplitudes refers to $S$-matrix elements, i.e., the on-shell amplitudes including external leg (wavefunction) corrections. A simple derivation of  $S$-matrix invariance is given in Ref.~\cite{Manohar:2018aog}, implementing field redefinitions as a change of variables in the functional integral.} Field redefinitions which do not include derivatives, such as $\phi(x) \to F(\phi(x))$ for a scalar field, can be viewed as a change of coordinates on the manifold where the scalar fields live, which does not change the dynamics of the theory. An example is chiral perturbation theory, where for two light flavors, the Goldstone boson manifold is the group $SU(2)$ which is isomorphic to the three-sphere $S^3$. Two common field choices are to use Cartesian coordinates $\bm{\tilde \pi}=(\tilde \pi^1, \tilde \pi^2,\tilde \pi^3,\tilde \pi^4)$ with the constraint $\bm{\tilde \pi \cdot \tilde \pi}=1$ for $S^3$, or the exponential parameterization $\exp ( i \bm{\pi \cdot \tau}/f)$ with $\bm{\pi}=(\pi^1,\pi^2,\pi^3)$ for the corresponding $SU(2)$ group element. The two forms lead to different off-shell correlation functions, but the same $S$-matrix elements.

The geometric approach was used to compute scattering amplitudes, and to characterize deviations from the Standard Model (SM) in terms of the curvature of the scalar manifold of the Higgs field~\cite{Alonso:2015fsp,Alonso:2016oah}. It was shown that deviations from the SM model for Higgs Effective Field Theory (HEFT) or Standard Model Effective Field Theory (SMEFT) have a simple universal form in terms of the curvature~\cite{Alonso:2015fsp,Alonso:2016oah}. Further work can be found in Refs.~\cite{Cohen:2021ucp,Alonso:2021rac}.
Recently, the geometric structure of scattering amplitudes under field redefinitions has been extended to include field redefinitions with derivatives and higher-spin fields through several approaches \cite{Cheung:2022vnd,Cohen:2022uuw}.

The geometric view of scattering amplitudes also has practical advantages. It reorganizes the calculation of amplitudes in terms of geometric invariants. Many terms in a Feynman-diagram expansion are organized into geometric quantities, leading to a more efficient calculation of the amplitude. It  provides a universal description of some scattering amplitudes---Higgs and longitudinal $W$ scattering in BSM models, soft scattering amplitudes for spontaneously broken theories~\cite{Alonso:2016oah,Cheung:2021yog}, and the renormalization group equations~\cite{Alonso:2015fsp,Alonso:2016oah,Alonso:2022ffe} in terms of the curvature.

The previous results were extended to include gauge fields and combine the scalar and gauge sectors in a unified framework~\cite{Helset:2022tlf}. Kinetic terms for the scalar and gauge fields are unified into a combined metric tensor with both scalar and gauge indices. The results unify scalar and gauge amplitudes, so that $\phi \phi \to \phi \phi$, $\phi \phi \to A A$ and $A A \to AA$ are different components of a single curvature tensor. Even though the starting metric is block-diagonal, with scalar and gauge components, the curvature tensor is not. Terms in the curvature tensor such as $\Gamma^{i}_{jr} \Gamma^{r}_{kl}$ have internal index sums which run over both scalar and gauge indices. They give terms in the scattering amplitude from diagrams with internal scalar and gauge exchange. The geometric analysis can be used to compute one-loop anomalous dimensions. We apply the methods in this paper to reproduce the renormalization group equations (RGEs) for the dimension-six even-parity bosonic operators in the SMEFT~\cite{Jenkins:2013zja,Jenkins:2013wua,Alonso:2013hga} as a check on the formalism. We then obtain the RGEs for dimension-eight even-parity bosonic operators in the SMEFT. Parts of the dimension-eight RGEs have been computed previously~\cite{Chala:2021pll,DasBakshi:2022mwk}, but a lot of terms are new. We agree with the previous results for the terms common to both calculations.

We will use the standard EFT power counting in $1/M$, where $M$ is a mass scale. Dimension-six contributions to the Lagrangian or RGE are proportional to $1/M^2$, dimension-eight contributions to $1/M^4$, etc. Section~\ref{sec:manifold} discusses the geometric formulation we use, including the combined scalar-gauge metric, covariant derivatives, and curvature. Section~\ref{sec:Renormalization} computes the second variation of the action using geodesic coordinates for the fluctuations, and the one-loop renormalization counterterms in terms of curvatures and field-strength tensors. The SMEFT Lagrangian to dimension eight, and the expressions for the metric and Killing vectors in the SMEFT are given in Sec.~\ref{sec:smeft}. The formalism of Secs.~\ref{sec:manifold} and \ref{sec:Renormalization} is applied to compute the RGEs in Sec.~\ref{sec:rge}. Operator counterterms have to be reduced to the canonical dimension-eight basis. These reduction expressions are lengthy, and given in App.~\ref{sec:Operators}, and the RGEs in the canonical basis are given in App.~\ref{sec:RGEresults}. Section~\ref{sec:Zeros} discusses the implications of our results for geometric zeros in the anomalous dimensions.
We conclude in Sec.~\ref{sec:Conclusion}.

\section{Field-Space Manifold}\label{sec:manifold}

Consider a theory of scalar and gauge bosons with interactions with at most two derivatives,\footnote{Higher-derivative interactions are linked to higher-derivative field redefinitions, which is outside the scope of this work. They have been considered in Refs.~\cite{Cheung:2022vnd,Cohen:2022uuw}.} and  ignore CP-violating interactions for simplicity. The general gauge-invariant Lagrangian takes the form
\begin{align}
	\label{eq:Lagr}
	\mathcal{L} &= 
	\frac{1}{2} h_{IJ}(\phi) (D_{\mu}\phi)^{I} (D^{\mu}\phi)^{J} - V(\phi)
	- \frac{1}{4} g_{AB}(\phi) F^{A}_{\mu\nu} F^{B\mu\nu}\,,
\end{align}
where $h_{IJ}(\phi)$, $V(\phi)$, and $g_{AB}(\phi)$ depend on the scalar fields, and
\begin{align}
	\label{eq:covD}
	(D_{\mu}\phi)^{I} &= \partial_{\mu} \phi^{I} + A^{B}_{\mu} t^{I}_{B}(\phi) , &
	F^{B}_{\mu\nu} &= \partial_{\mu} A^{B}_{\nu} - \partial_{\nu} A^{B}_{\mu} - \f{B}{C}{D} A^{C}_{\mu} A^{D}_{\nu} ,
\end{align}
where $t^I_A(\phi)$ are Killing vectors of the scalar manifold, so they generate a symmetry.\footnote{More details can be found in Ref.~\cite{Alonso:2016oah}.} The Lie derivative of the scalar metric $h_{IJ}$ vanishes,
\begin{align}
	\label{eq:LieMetricH}
\left( \mathcal{L}_{t_A} h \right)_{IJ} =	t^{K}_{A} h_{IJ,K} + h_{KJ} t^{K}_{A,I} + h_{IK} t^{K}_{A,J} = 0 ,
\end{align}
where $h_{IJ,K} = \partial_{K} h_{IJ}$ and $t^{I}_{A,J} = \partial_{J} t^{I}_{A}$.  The Killing vectors satisfy the Lie bracket relations
\begin{align}
[t_A,t_B]^I =& \f CAB t^I_C \,,
\end{align}
and the relation
\begin{align}\label{2.5}
\nabla_J t_{IA} = - \nabla_I t_{JA} \,,
\end{align}
where $t_{IA} = h_{IJ} t^J_A$.
The gauge coupling constant is included in $t^I_A$, and hence also in the structure constants $\f CAB$.
%
%
%
%

The kinetic term coefficient for the scalars, $h_{IJ}(\phi)$, can be interpreted as a metric in scalar field space~\cite{Meetz:1969as}, and transforms as a metric under field redefinitions. The kinetic term coefficient for the gauge fields, $g_{AB}(\phi)$, which depends on the scalars, is symmetric under $A \leftrightarrow B$, and transforms as an invariant tensor with two adjoint indices under action by the Killing vector $t^I_A$,
\begin{align}
\label{eq:LieMetricG}
g_{AB,I} \, t^I_C - \f DCA\ g_{DB} - \f DCB\ g_{AD} =0 \,.
\end{align}
We extend the notion of a field-space manifold to include  gauge fields, where $g_{AB}(\phi)$ will take center stage, and unify the scalar and gauge sectors, so Eqs.~\eqref{eq:LieMetricH} and~\eqref{eq:LieMetricG} are components of a single equation. This  also provides a unified description of scalar and gauge amplitudes.

We group the scalars and gauge bosons into real multiplets $\phi^{I}$ and $A^{B\mu_B}$, where $I,J,K,\dots$ are scalar indices and $(A\mu_{A}), (B\mu_{B}),\dots$ are gauge and Lorentz indices, treated as a combined index.  We will use $i,j,k,\ldots$ to run over both scalar and gauge indices. We define a combined metric
\begin{align}
	\label{eq:fullMetric}
	\widetilde g_{ij} = \begin{pmatrix}
		h_{IJ} & 0 \\
		0 & -\eta_{\mu_A \mu_B} \; g_{AB}
	\end{pmatrix} 
\end{align}
from the scalar and gauge kinetic terms. The quadratic part of the gauge kinetic term is
\begin{align}
	\mathcal{L} &= - \frac{1}{2} g_{AB}(\phi) \left[(\partial_\mu A^A_\nu)(\partial_\mu A^B_\nu) - (\partial_\mu A^A_\mu)(\partial_\nu A^B_\nu) \right] \,.
\end{align}
The first term in the square brackets motivates the choice in Eq.~\eqref{eq:fullMetric}. The second term is cancelled by the gauge-fixing term.

In earlier works~\cite{Alonso:2015fsp,Alonso:2016oah}, the metric used was the scalar metric $h_{IJ}$. This metric gives the Christoffel symbol
\begin{align}
	\Gamma^{I}_{JK} = \frac{1}{2} h^{IL}\left(h_{JL,K} + h_{LK,J} - h_{JK,L} \right) ,
	\label{2.11}
\end{align}
and Riemann curvature
\begin{align}
	R_{IJKL} = h_{IM} \left(\partial_{K} \Gamma^{M}_{LJ} - \partial_{L} \Gamma^{M}_{KJ} + \Gamma^{M}_{KN} \Gamma^{N}_{LJ} - \Gamma^{M}_{LN} \Gamma^{N}_{KJ} \right) .
\end{align}
Covariant derivatives using the connection in Eq.~\eqref{2.11} are denoted by $\nabla_I$, where only the scalar indices are treated as active indices. We can compare these with quantities derived from the metric in Eq.~\eqref{eq:fullMetric}, which we denote with a tilde superscript. The various components of the Christoffel symbol $\widetilde \Gamma^i_{jk}$ are
\begin{subequations}\label{eq:Christoffel}
\begin{align}
	\label{eq:ChristoffelIJK}
	\widetilde \Gamma^{I}_{JK} &= \Gamma^{I}_{JK} , \\[5pt]
	\label{eq:ChristoffelAJK}
	\widetilde \Gamma^{(A\mu_A)}_{JK} &=  \widetilde \Gamma^{I}_{(A\mu_A)K} = \widetilde \Gamma^{(C\mu_C)}_{(A\mu_A)(B\mu_B)} = 0 , \\[5pt]
	\label{eq:ChristoffelIAB}
	\widetilde \Gamma^{I}_{(A\mu_A)(B\mu_B)} &= \frac{1}{2} h^{IJ} \nabla_{J} g_{AB} \eta_{\mu_{A} \mu_{B}} , \\[5pt]
	\label{eq:ChristoffelAIB}
\widetilde \Gamma^{(A\mu_A)}_{I(B\mu_B)} &= \frac{1}{2} g^{AC} \nabla_{I} g_{CB} \delta^{\mu_{A}}_{ \mu_{B}} ,
\end{align}
\end{subequations}
where $\nabla_{I} g_{AB} = g_{AB,I}$ is the covariant derivative using the connection $\nabla_I$, since $A,B$ are not active indices for $\nabla_I$. Christoffel symbols with an odd number of gauge indices vanish. We will also use the notation
\begin{subequations}\label{2.13}
\begin{align}
	\widetilde \Gamma^{I}_{(A\mu_A)(B\mu_B)} &\equiv \Gamma^I_{AB} (-\eta_{\mu_A \mu_B}) \,, &
	\Gamma^I_{AB}  = & - \frac{1}{2} h^{IJ} \nabla_{J} g_{AB} , \\[5pt]
\widetilde \Gamma^{(A\mu_A)}_{I(B\mu_B)} &\equiv \Gamma^A_{IB} \delta^{\mu_A}_{\mu_B}\,, &
 \Gamma^A_{IB} = & \frac{1}{2} g^{AC} \nabla_{I} g_{CB} ,
\end{align}
\end{subequations}
where $\eta_{\mu_A \mu_B}$ and $ \delta^{\mu_A}_{\mu_B}$  have been factored out. Even though the metric in Eq.~\eqref{eq:fullMetric} is block diagonal, we get non-zero mixed Christoffel symbols with both scalar and gauge indices.

The Riemann curvature tensor components $\widetilde R_{ijkl}$ are computed from the Christoffel symbols $\widetilde \Gamma^i_{jk}$, and the summation over indices runs over both scalar and gauge indices. The components of $\widetilde R_{ijkl}$ are
\begin{subequations}\label{eq:Riemann}
\begin{align}
	\widetilde R_{IJKL} =& R_{IJKL} , \\[5pt]
	\widetilde R_{(A\mu_A)JKL} =& R_{(A\mu_A)(B\mu_B)(C\mu_C)L} = 0 , \\[5pt]
	\widetilde R_{IJ(A\mu_A) (B\mu_B) } =& \left( \frac14 (\nabla_I g_{AC}) g^{CD} (\nabla_J g_{BD} ) -
	\frac14  (\nabla_J g_{AC}) g^{CD} (\nabla_I g_{BD} ) \right) \eta_{\mu_A \mu_B} , \\[5pt]
\widetilde R_{I(A\mu_A) J (B\mu_B) } =&  \left(  \frac{1}{2} \nabla_{I} \nabla_{J} g_{AB} - \frac{1}{4} (\nabla_{J} g_{AC}) g^{CD} (\nabla_{I} g_{BD}) \right) \eta_{\mu_A \mu_B} , \\[5pt]
	\widetilde R_{(A\mu_A) (B\mu_B) (C\mu_C) (D\mu_D)} =& - \frac{1}{4} (\nabla_{I} g_{AC}) h^{IJ}( \nabla_{J} g_{BD}) \; \eta_{\mu_A \mu_C} \eta_{\mu_B \mu_D}
	\nonumber \\ &
	+ \frac{1}{4} (\nabla_{I} g_{AD} ) h^{IJ} ( \nabla_{J} g_{BC}) \; \eta_{\mu_A \mu_D} \eta_{\mu_B \mu_C} .
\end{align}
\end{subequations}
Curvature components with an odd number of gauge indices vanish.  Here
\begin{align}
\nabla_{I} \nabla_{J} g_{AB} =  g_{AB,IJ} - \Gamma^K_{IJ} g_{AB,K} \,,
\end{align}
since only the scalar indices are active indices for $\nabla_I$. The gauge curvature obeys the Bianchi identities
\begin{align}
	\widetilde  R_{(A\mu_{A})(B\mu_{B}) IJ} + \widetilde R_{(A\mu_{A}) IJ(B\mu_{B})} + \widetilde R_{(A\mu_{A})J (B\mu_{B}) I}  = 0 \,,
\end{align}
and similarly for $\widetilde R_{IJKL}$ and $\widetilde R_{(A\mu_A) (B\mu_B) (C\mu_C) (D\mu_D)}$.

We will also use the covariant derivative $\widetilde \nabla_I$ using the Christoffel connection $\widetilde \Gamma^i_{jk}$ in Eq.~\eqref{eq:Christoffel}, where scalar and gauge indices are both active indices. One quantity which enters in helicity amplitudes is \cite{Helset:2022tlf}
\begin{align}
	\widetilde \nabla_{I} \nabla_{J} \left(  g_{AB} \eta_{\mu_A \mu_B} \right) =& \left( \nabla_{I} \nabla_{J} g_{AB} - \frac{1}{2} \nabla_{I} g_{AC} \; g^{CD} \; \nabla_{J} g_{BD} -  \frac{1}{2} \nabla_{J}  g_{AC} \; g^{CD} \; \nabla_{I} g_{BD} \right) \eta_{\mu_A \mu_B} \,,
\end{align}
where $A,B$ are active indices for the combined covariant derivative $\widetilde \nabla$, but not for the scalar covariant derivative $\nabla$. As in Eq.~\eqref{2.13}, it is convenient to factor out $\eta_{\mu_A\mu_B}$ from both sides,
\begin{align}\label{2.19}
	\widetilde \nabla_{I} \nabla_{J} g_{AB} \equiv & \left( \nabla_{I} \nabla_{J} g_{AB} - \frac{1}{2} \nabla_{I} g_{AC} \; g^{CD} \; \nabla_{J} g_{BD} -  \frac{1}{2} \nabla_{J}  g_{AC} \; g^{CD} \; \nabla_{I} g_{BD} \right) =\widetilde \nabla_{J} \nabla_{I} g_{AB} \, .
\end{align}
These geometric quantities arise in the calculation of the renormalization group equations.

\section{Renormalization}\label{sec:Renormalization}

\begin{figure}
\begin{center}
\begin{tikzpicture}
\draw (0,0) circle (1);
\filldraw (90:1) circle (0.05);
\filldraw (210:1) circle (0.05);
\filldraw (-30:1) circle (0.05);
\draw (90:1) -- +(90:1);
\draw (90:1) -- +(60:1);
\draw (90:1) -- +(120:1);
\draw (-30:1) -- +(-60:1);
\draw (-30:1) -- +(0:1);
\draw (210:1) -- +(210:1);
\end{tikzpicture}
\end{center}
\caption{\label{loop} Generic one-loop graph. The internal lines are fluctuation fields, $\eta^I$ and $\zeta^A_\mu$, and the external lines are the background fields $\Phi^I$ and $\mathsf{A}^{B\mu_{B}}$. All interaction vertices are quadratic in the fluctuations.}
\end{figure}
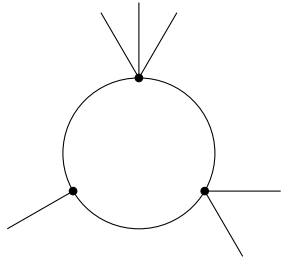

The one-loop renormalization of the Lagrangian in Eq.~\eqref{eq:Lagr} can be computed using the background field method. The scalar fields are written as the sum of a background field $\Phi$ plus fluctuation $\eta$, $\phi^I \to \Phi^I + \eta^I$. The one-loop renormalization is computed by expanding the Lagrangian to second order in the fluctuations, and then integrating over the fluctuations. 
A generic one-loop graph that contributes to the RGEs is shown in Fig.~\ref{loop}.

The expansion $\phi^I \to \Phi^I + \eta^I$ is not a covariant expansion to second order in the fluctuation, and it is better to use instead an expansion in geodesic coordinates~\cite{Honerkamp:1971xtx,Honerkamp:1971sh,Alonso:2015fsp,Alonso:2016oah}
\begin{align}
	\phi^{I} = \Phi^{I} + \eta^{I} - \frac{1}{2} \Gamma^{I}_{JK} \eta^{J}\eta^{K} + \dots 
\end{align}
This results in a covariant second variation of the action.  In our case, we use geodesic coordinates for both scalar and gauge fluctuations,
\begin{align}
	\eta^{i} = \begin{pmatrix}
		\eta^{I} \\
		\zeta^{A\mu_A}
	\end{pmatrix} \,,
\end{align}
with the connection derived from the combined metric in Eq.~\eqref{eq:fullMetric}. The expansions of the fields are
\begin{subequations}\label{4.3}
\begin{align}
	\phi^{I} &= \Phi^{I} + \eta^{I} - \frac{1}{2} \widetilde \Gamma^{I}_{jk} \eta^{j} \eta^{k}+ \dots  \nn
	&=\Phi^{I} + \eta^{I} - \frac{1}{2} \widetilde \Gamma^{I}_{JK} \eta^{J} \eta^{K} - \frac{1}{2} \widetilde \Gamma^{I}_{(A \mu_A)(B \mu_B)} \zeta^{A \mu_A} \zeta^{B \mu_B}+ \dots \nn
	&=\Phi^{I} + \eta^{I} - \frac{1}{2}  \Gamma^{I}_{JK} \eta^{J} \eta^{K} + \frac{1}{2}  \Gamma^{I}_{AB} \zeta^{A \mu} \zeta^B_\mu+ \dots \,, \\
	A^{B\mu_{B}} &=  \mathsf{A}^{B\mu_{B}} + \zeta^{B\mu_B} - \frac{1}{2} \widetilde \Gamma^{(B\mu_B)}_{jk} \eta^{j} \eta^{k} + \dots \nn
	& = \mathsf{A}^{B\mu_{B}} + \zeta^{B\mu_B} - \frac{1}{2} \widetilde \Gamma^{(B\mu_B)}_{(C \mu_C) K} \zeta^{C \mu_C} \eta^{K} - \frac{1}{2} \widetilde \Gamma^{(B\mu_B)}_{J (C \mu_C)} \eta^{J} \zeta^{C \mu_C} + \dots \nn
	& = \mathsf{A}^{B\mu_{B}} + \zeta^{B\mu_B} -  \Gamma^{B}_{CK} \zeta^{C \mu_B} \eta^{K} 
 + \dots \,,
 \end{align}
\end{subequations}
where $\Phi^{I} $ and $\mathsf{A}^{B \mu_B}$ are the background fields. After expanding the action, we will simply use $\phi^{I}$ and $A^{B\mu_B}$ for the background fields when there is no ambiguity. Note that with the choice in Eq.~\eqref{4.3},  there is mixing between the scalar and gauge fluctuations at second order. 

The computation of the variation of the action to second order is a lengthy calculation. The substitution in Eq.~\eqref{4.3} is used for the fields, and then the action is expanded to second order in $\eta$ and $\zeta$. There is considerable simplification when using the expansion in Eq.~\eqref{4.3} and the symmetries in Eqs.~\eqref{eq:LieMetricH} and~\eqref{eq:LieMetricG}. Using geodesic fluctuations gives the resultant fluctuations in terms of geometric quantities.  In computing variations of the action, it is useful to define various covariant derivatives. The  gauge covariant derivative of $\phi$ is
\begin{align}
	(D_\mu \phi)^I &= \partial_\mu \phi^I + A^B_\mu t^I_B(\phi) \,,
\end{align}
and the gauge covariant derivative of a gauge adjoint such as $F^A_{\alpha \beta}$ or $\zeta^{A \alpha}$ is
\begin{align}
(D_\mu F_{\alpha \beta})^A &= \partial_\mu F^A_{\alpha \beta} -\f A B C A^B_\mu F^C_{\alpha \beta} \,, &
(D_\mu \zeta^{\alpha})^{A } &= \partial_\mu \zeta^{A\alpha }-\f A B C A^B_\mu \zeta^{C\alpha } \,.
\end{align}
The derivative of the scalar fluctuation that is covariant w.r.t.\ gauge transformations and scalar manifold coordinate transformations is~\cite{Alonso:2016oah}
\begin{align}
(\newD_\mu \eta)^I &= \partial_\mu \eta^I + t^I_{B,K} A^B_\mu \eta^K + \Gamma^{I}_{JK} (D_{\mu} \phi)^{J} \eta^{K} \,.
\end{align}
Since $(D_\mu \phi)^I$ transforms like the fluctuation $\eta^I$, the covariant second derivative of $\phi$ is
\begin{align}
(\mathscr{D}_\nu D_\mu  \phi)^I &= \partial_\nu (D_\mu  \phi)^I  + t^I_{B,J} A_\nu^B (D_\mu \phi)^J + \Gamma^I_{KL} (D_\nu \phi)^K   (D_\mu  \phi)^L \,.
\end{align}
Generalizing to the combined metric in Eq.~\eqref{eq:fullMetric}, we define a covariant derivative $\nnewD$ w.r.t.\ both the gauge field and the background metric $\widetilde g_{ij}$ analogous to the definition of $\newD$ in Ref.~\cite{Alonso:2016oah}. Let
\begin{align}
	Z_\mu^i =& \begin{bmatrix} (D_\mu \phi)^I \\ F_{\mu}^{A \; \mu_A} \end{bmatrix}
\end{align}
be the analog of $D_\mu \phi$ in the combined scalar-gauge space, and define
\begin{align}
( \nnewD_\mu \eta)^I &= \partial_\mu \eta^I + t^I_{B,K} A^B_\mu \eta^K + \widetilde \Gamma^{I}_{jk}Z_\mu^j \eta^{k} \nn
 &= \partial_\mu \eta^I + t^I_{B,K} A^B_\mu \eta^K + \Gamma^{I}_{JK } (D_\mu \phi)^J \eta^{K}
 +\widetilde \Gamma^{I}_{(A\mu_A)(B \mu_B)} F_{\mu}^{A \; \mu_A} \zeta^{B\mu_B} \nn
&= \partial_\mu \eta^I + t^I_{B,K} A^B_\mu \eta^K + \Gamma^{I}_{JK} (D_{\mu} \phi)^{J} \eta^{K}
- \Gamma^{I}_{AB} F^A_{\mu \nu} \zeta^{B\nu} \nn
&= ( \newD_\mu \eta)^I  - \Gamma^{I}_{AB} F^A_{\mu \nu} \zeta^{B\nu} \,,
\end{align}
and similarly
\begin{align}\label{4.10}
\left(\nnewD_{\mu} \zeta \right)^{(A \mu_A)} &= \partial_\mu \zeta^{A \mu_A} - \f A C D A^C_\mu \zeta^{D \mu_A} + \widetilde \Gamma^{(A\mu_A)}_{ij} Z^i_\mu \eta^j \nn
					     &= \partial_\mu \zeta^{A \mu_A} - \f A C D A^C_\mu \zeta^{D \mu_A} + \widetilde \Gamma^{(A\mu_A)}_{I(B \mu_B)}(D_\mu \phi)^I \zeta^{B \mu_B} + \widetilde \Gamma^{(A\mu_A)}_{(B \mu_B) I} F_{\mu}^{B \; \mu_B} \eta^I \nn
					     &= \left(D_{\mu} \zeta^{\mu_A} \right)^{A} + \Gamma^{A}_{IB}(D_\mu \phi)^I \zeta^{B \mu_A} +  \Gamma^{A}_{B I} F_{\mu}^{B \; \mu_A} \eta^I  \,.
\end{align}
The final expressions for the one-loop counterterms simplify greatly when written in terms of $\nnewD_\mu \eta$ and $\nnewD_\mu \zeta$.

\subsection{First Order Variation}

The first variation of the action in Eq.~\eqref{eq:Lagr} is
\begin{align}
	\delta_\eta S &= \int d^{4} x \biggl\{ - h_{IJ}  (\mathscr{D}_\mu D^{\mu} \phi)^{J} 
	- \frac{1}{4} g_{AB,I} F^{A}_{\mu\nu} F^{B\mu\nu}  - \nabla_I V  \biggr\} \eta^I 
	\label{31}
\end{align}
under scalar fluctuations, and
\begin{align}
	\delta_\zeta S &= \int d^{4} x \biggl\{  h_{IJ}   t^I_B  (D_{\nu} \phi)^{J}  + g_{AB,I} (D^\mu \phi)^I F_{\mu \nu}^A + g_{AB} (D^\mu F_{\mu \nu})^A   \biggr\} \zeta^{B\nu}
\end{align}
under gauge fluctuations, and $\delta_\eta S=0$, $\delta_\zeta S=0$ are the classical equations of motion.

\subsection{Second Order Variation}

Obtaining the second order variation of the action is a tedious computation, with many terms, which collapse into a covariant expression when using the symmetry conditions
in Eqs.~\eqref{eq:LieMetricH} and~\eqref{eq:LieMetricG}.  The second order terms can be divided into the scalar variation $\delta_{\eta\eta}$, the gauge variation $\delta_{\zeta \zeta}$, and the mixed variation $\delta_{\eta \zeta}$. We have used the geodesic fluctuations in Eq.~\eqref{4.3} to compute the second order variation, which results in a covariant expression and simplifies the final result. To the second variation, we have added a gauge-fixing term, Eq.~\eqref{4.21}, to eliminate terms linear in $(\newD_\mu \zeta^\mu)^A$. The gauge-fixing term  is included in the expressions below.

\vspace{0.25cm}

\noindent\framebox{$\eta \eta$:} The scalar variation is
\begin{align}
	\delta_{\eta\eta} S &= \frac{1}{2} \int d^{4} x\  \biggl\{
		 h_{IJ} \left(\nnewD_{\mu} \eta \right)^{I}  \left(\nnewD_{\mu} \eta \right)^{J}  + 
	\Bigl[  - \widetilde R_{IKJL} (D_{\mu} \phi)^{K} (D^{\mu} \phi)^{L} -  (\nabla_{I}\nabla_{J}V  )   \nn
&		 -\frac{1}{4}  \left(   \nabla_I \nabla_J g_{AB} - \Gamma^C_{I A} g_{CB,J}- \Gamma^C_{I B} g_{AC,J}  \right) F^{A\mu\nu} F^B_{\mu\nu}	 -     h_{IK} h_{JL} g^{AB}  t^K_A  t^L_B \Bigr] \eta^I \eta^J 	  \biggr\} \,,
\end{align}
which can be written in the more compact form
\begin{align}\label{4.15}
	\delta_{\eta\eta} S &= \frac{1}{2} \int d^{4} x  \ \biggl\{
		 h_{IJ} \left(\nnewD_{\mu} \eta \right)^{I}  \left(\nnewD_{\mu} \eta \right)^{J}  + 
	\Bigl[  - \widetilde R_{IKJL} (D_{\mu} \phi)^{K} (D^{\mu} \phi)^{L} -  (\nabla_{I}\nabla_{J}V  )   \nn
&\qquad \qquad\qquad\qquad\qquad\qquad\qquad\qquad\quad		 -\frac{1}{4}  \left(  \widetilde \nabla_I \nabla_J g_{AB}  \right) F^{A\mu\nu} F^B_{\mu\nu}	 -     t_{IA}  t^B_J \Bigr] \eta^I \eta^J 	  \biggr\} \,.
\end{align}
The covariant derivative $\widetilde \nabla_I \nabla_J g_{AB}$ is given in Eq.~\eqref{2.19}. Scalar and gauge indices on the Killing vector $t^I_A$ are lowered and raised by the metrics $h_{IJ}$ and $g_{AB}$ and their inverses,
\begin{align}
t_{IA} = & h_{IJ} t^J_A \,, &
t^A_I =& g^{AB} h_{IJ} t^J_B\,.
\end{align}

\vspace{0.25cm}
\noindent\framebox{$\eta \zeta$:} The mixed variation is
\begin{align}
	\delta_{\eta\zeta} S =
	\int d^{4}x &
	\biggl[ { \left( h_{KJ} \nabla_I t^J_{A}  - h_{IJ} \nabla_K t^J_{A} \right)(D_\mu \phi)^K}_{}  - { \frac12(\nabla_J \nabla_I g_{AB}) (D^\nu \phi)^J F_{\mu \nu}^B }   \nn
	   &  -\frac12  g^{BD} g_{AD,I}  h_{LJ}   t^L_B  (D_{\mu} \phi)^{J}   + \frac12  g^{BD} g_{AD,I}  g_{CB,L} (D^\nu \phi)^L F_{\mu \nu}^C  \nn
	   	    & {+ g_{BA,K}( D_\mu  \phi)^K    h_{IL} g^{BG} t^L_G}_{}  
	    { - \frac14 (D^\nu \phi)^J g_{AB,J}  g^{BD} g_{DC,I} F_{\mu \nu}^C} \biggr] \eta^I \zeta^{A\mu} \,.
\end{align}
The first term can be rewritten using the identity
\begin{align}
 h_{KJ} \nabla_I t^J_{A}  - h_{IJ} \nabla_K t^J_{A} =& 2 h_{KJ} \nabla_I t^J_{A}
 = 2 \nabla_{I} t_{KA} \,,
\end{align}
which follows from Eq.~\eqref{2.5} since $t_A$ is a Killing vector. The entire expression reduces to
\begin{align}\label{4.18}
	\delta_{\eta\zeta} S &= \int d^{4}x\ \biggl\{ \left[ 2 (\nabla_{I}t_{JA}) + t^{B}_{I} (\nabla_{J} g_{AB}) - \frac{1}{2} t^{B}_{J} (\nabla_{I} g_{AB}) \right] (D_{\mu_A}\phi)^{J} 
		\nn &\qquad\qquad\quad 
		 + \left(- \widetilde R_{(A\mu_A) I(B\mu_B)J} +2 \widetilde R_{IJ(A\mu_A)(B\mu_B)} \right) F^{B\mu_B \rho} (D_{\rho} \phi)^{J} 
	\biggr\} \eta^{I} \zeta^{A\mu_A}  
\end{align}
using the combined curvature $\widetilde R$ defined in Eq.~\eqref{eq:Riemann}.

\vspace{0.25cm}
\noindent\framebox{$\zeta \zeta$:} The gauge variation is
\begin{align}\label{4.20}
	\delta_{\zeta\zeta}  S	 & =  \frac12 \int \rd^4x \ \bigg\{ - g_{AB}\, \eta_{\mu_A\mu_B}  (\nnewD_\mu \zeta)^{A\mu_A} (\nnewD^\mu \zeta)^{B\mu_B} 
		\nn &
		+ \Bigl[ t_{IA}t^I_B \eta_{\mu \nu} 
 -\widetilde R_{I(A\mu)J(B\nu)} (D_\alpha \phi)^I   (D^\alpha \phi)^J     + \frac12  g_{AB,I} \left( ( \nnewD_\mu D_\nu  \phi)^I + ( \nnewD_\nu D_\mu  \phi)^I \right)   \nn
	 &  + \left( \nabla_I \nabla_J g_{AB}   -  g_{AD,I}  g^{DG}g_{GB,J} \right) (D_\mu \phi)^I (D_\nu \phi)^J    \nn
	 	 &  + \frac{1}{2}  \left( g_{DB}  \f DCA -g_{DA}  \f DCB  +2 g_{CD}  \f DAB \right)F^{C}_{\mu\nu}   
	 - \frac14  g_{DB,K}  h^{KL} g_{CA,L} F_{\alpha \mu}^C   F_{\alpha  \nu}^D \nn
&
+ \frac{1}{8}  h^{IM} g_{AB,M} g_{CD,I} F^{C}_{\alpha \beta} F^{D\alpha \beta}  \eta_{\mu \nu}  +\frac12   h^{IM} g_{AB,M} V_{,I} 
\eta_{\mu \nu}    
 \Bigr] \zeta^{A\mu}  \zeta^{B\nu} \biggr\} \,.
\end{align}

\vspace{0.25cm}

\noindent\framebox{Gauge-fixing term:} The gauge-fixing term, which has been included in the above second variation of the action, is
\begin{align}\label{4.21}
S_{\text{g.f.}} &=	 - \frac12 \int \rd^4 x \ g_{AB}  \mathcal{G}^A \mathcal{G}^B \,, \nn
\mathcal{G}^A &= (\nnewD_\mu \zeta)^{A\mu} + \frac12 g^{AC} g_{CB,I}( D_\mu  \phi)^I    \zeta^{B\mu} - h_{IJ} g^{AB} t^J_B \eta^I  \nn
	      &= (\nnewD_\mu \zeta)^{A\mu} + \Gamma^A_{IB} ( D_\mu  \phi)^I    \zeta^{B\mu} - t^A_I \eta^I \,.
\end{align}
This is an extension of the gauge-fixing term in Ref.~\cite{Helset:2018fgq}.
The gauge-fixing term in Eq.~\eqref{4.21} has been chosen to eliminate terms linear in $(\nnewD_\mu \zeta)^{A\mu}$ in the second variation of the action, and to make the $\zeta$ kinetic term invertible. Physical results do not depend on the choice of gauge-fixing term.

\vspace{0.25cm}
\noindent\framebox{Ghosts:} There is also a ghost Lagrangian which depends on the gauge variation of the gauge-fixing term in Eq.~\eqref{4.21}. Under a gauge transformation with parameter $\theta^A$,
\begin{align}
\delta \eta^I &= t^I_{A,J} \eta^J \theta^A \,, &
\delta \zeta^A_\mu &= - \partial_\mu \theta^A -\f ABC \theta^B ( \mathsf{A}_\mu^C + \zeta_\mu^C)\,,
\end{align}
the ghost Lagrangian takes the form
\begin{align}
S_{\text{ghost}} =& \int \rd^4x \ \overline c_A \frac{\delta \mathcal{G}^A}{\delta \theta^B} c^B \nn
=&  \int \rd^4x \ 
\biggl\{ (D_\mu \overline c)_A (D^\mu c)^B + (D_\mu \overline c)_A    \f ABC  \zeta_\mu^C c^B - 2 \overline c_A \Gamma^{A}_{IB} (D_{\mu} \phi)^{I} (D_\mu c)^B \nn
& - 2 \overline c_A \Gamma^{A}_{IB} (D_{\mu} \phi)^{I}  \f ABC \zeta_\mu^C c^B  - \overline c_A h_{IJ} g^{AC} t^J_Ct^I_{B}  c^B - \overline c_A h_{IJ} g^{AC} t^J_Ct^I_{B,K} \eta^K c^B \biggr\}\,,
\end{align}
where $c$ and $\overline c$ are the anticommuting ghost and anti-ghost fields. The covariant derivative of the ghost and anti-ghost analogous to Eq.~\eqref{4.10} is
\begin{align}
\left(\nnewD_{\mu} c \right)^{A} &= \partial_\mu c^A - \f A BC A^B_\mu c^C  + \Gamma^{A}_{IB} (D_{\mu} \phi)^{I} c^{B}  
= (D_{\mu} c)^{A} + \Gamma^{A}_{IB} (D_{\mu} \phi)^{I} c^{B} \,, \nn
\left(\nnewD_{\mu} \overline c \right)_{A} &= \partial_\mu \overline c_A - \overline c_C \f C BA A^B_\mu   +\overline c_B \Gamma^{B}_{IA} (D_{\mu} \phi)^{I}   
= (D_{\mu} \overline c)_{A} + \overline c_B \Gamma^{B}_{IA} (D_{\mu} \phi)^{I}  \,,
\end{align}
in terms of which the ghost action is
\begin{align}\label{4.26}
	& S_{\rm ghost} =  \int d^4 x \biggl\{ (\nnewD_{\mu} \overline c)_{A} (\nnewD^{\mu} c)^{A} 
		\nn &
	+ \overline c_{A} \Bigl[ \left(\frac12 g^{AE} \nabla_I \nabla_J g_{EB} - \frac14 g^{AE} g_{EC,I} g^{CD} g_{DB,J} \right) (D_{\mu} \phi)^{I} (D^{\mu} \phi)^{J} +   \Gamma^{A}_{IB} (\newD^\mu D_\mu \phi)^{I} - t^{IA} t_{IB} \Bigr] c^{B}  \nn
	&+ (D_\mu \overline c)_A\f A B C \zeta^C_\mu c^B - 2 \overline c_A \Gamma^A_{IB} (D_\mu \phi)^I \f A B C \zeta_\mu^C c^B -\overline c_A h_{IJ} g^{AC} t^J_C t^I_{B,K} \eta^K c^B \biggr\}  \,,
\end{align}
which can be written in the simpler form
\begin{align}\label{4.26a}
	S_{\rm ghost} =&  \int d^4 x \biggl\{ (\nnewD_{\mu} \overline c)_{A} (\nnewD^{\mu} c)^{A}  + \overline c_{A} \Bigl[g^{AC} \frac12 \widetilde \nabla_I \nabla_J g_{CB}  (D_{\mu} \phi)^{I} (D^{\mu} \phi)^{J} 
	+   \Gamma^{A}_{IB} (\newD^\mu D_\mu \phi)^{I} - t^{IA} t_{IB} \Bigr] c^{B}  \nn
	&+ \left[ (D_\mu \overline c)_A\f A B C \zeta^C_\mu  - 2 \overline c_A \Gamma^A_{IB} (D_\mu \phi)^I \f A B C \zeta_\mu^C  -\overline c_A  t^A_I t^I_{B,K} \eta^K \right] c^B \biggr\}  \,.
\end{align}
The last line of the ghost action is cubic in the fluctuation fields, and not needed for the one-loop functional integral over fluctuations.

\subsection{One-Loop Counterterms}

The divergent one-loop contributions are calculated from the second variation of the action. The general form was  first computed in Ref.~\cite{tHooft:1973bhk} and extended to a kinetic term with non-trivial metric in Ref.~\cite{Alonso:2016oah}. In the purely scalar case, if the second variation has the form
\begin{align}
\delta_{\eta \eta} S =& \frac12 \int \rd^4x \ \biggl\{ h_{IJ} (\newD_\mu \eta)^I (\newD_\mu \eta)^J 
+  X_{IJ} \eta^I \eta^J \biggr \} \,,
\end{align}
then the infinite part of the one-loop functional integral in $4-2\epsilon$ dimensions is
\begin{align}
	\label{eq:divergence}
	\Delta S =& \frac{1}{32 \pi^2 \epsilon}   \int \rd^4x \ \left\{ \frac{1}{12} {\rm Tr} \left[ Y_{\mu\nu} Y^{\mu\nu} \right] + \frac{1}{2} {\rm Tr} \left[ \mathcal{X}^2 \right] \right\} \,,
\end{align}
where
\begin{align}
	\left[ Y_{\mu\nu} \right]^{I}_{\; J} =& \left[ \newD_{\mu} , \newD_{\nu} \right]^{I}_{\; J} \,, &
	\mathcal{X}^I{}_J =& h^{IK} X_{KJ} \,.
\end{align}

In our case, we can use the above results treating $X$ and $Y$ as matrices in the combined scalar-gauge space, and subtract the corresponding expression for the ghosts. The components of $\mathcal{X}$,
\begin{align}
\mathcal{X} =& \begin{bmatrix} [\mathcal{X}_{\eta\eta}]^I{}_J  & [\mathcal{X}_{\eta\zeta}]^I{}_{(B \mu_B)} \\[5pt]
[\mathcal{X}_{\eta\zeta}]^{(A\mu_A)} {}_J  & [\mathcal{X}_{\zeta\zeta}]^{(A \mu_A)} {}_{(B \mu_B)}  
\end{bmatrix} \,,
\end{align}
can be read off from Eqs.~\eqref{4.15},~\eqref{4.18}, and~\eqref{4.20}, and $\mathcal{X}^A{}_B$ for the ghosts from Eq.~\eqref{4.26}.

The covariant derivative in the combined scalar-gauge space is
\begin{align}
\nnewD_{\mu} \begin{bmatrix} \eta^I \\ \zeta_\lambda^A \end{bmatrix} &= \partial_\mu  \begin{bmatrix} \eta^I \\ \zeta_\lambda^A \end{bmatrix} 
+ \begin{bmatrix}  t^I_{C,J} A^C_\mu  + \Gamma^{I}_{LJ} (D_{\mu} \phi)^{L} & - \Gamma^I_{CB} F_{\mu \sigma}^C  \\   \Gamma^{A}_{CJ} F^{C}_{\mu\lambda}  &  - \f ACB A_\mu^C \eta_{\lambda \sigma} + \Gamma^{A}_{LB} (D_{\mu} \phi)^{L}  \eta_{\lambda \sigma }\end{bmatrix}  \begin{bmatrix} \eta^J \\ \zeta_\sigma^B \end{bmatrix} \,,
\label{17}
\end{align}
and the commutator of  covariant derivatives $\nnewD$ takes a very simple form, 
\begin{align}
 \left[ \nnewD_{\mu} , \nnewD_{\nu} \right]^{i}_{\;\; j}= \left[ \widetilde Y_{\mu\nu} \right]^{i}_{\;\; j} = \widetilde R^{i}_{\; jkl} (D_{\mu} \phiF)^{k} (D_{\nu} \phiF)^{l} + \widetilde \nabla_{j} \widetilde t^{i}_{C} F^{C}_{\mu\nu} \,,
\end{align}
extending Ref.~\cite[(3.45)]{Alonso:2016oah},
where the combined Killing vector is
\begin{align}
	\widetilde t^{i}_{B} = \begin{bmatrix}
		t^{I}_{B} \\
		- \delta^{A}_{B} \partial_{\mu_A} + \f {A}{C}{B} A^{C}_{\mu_A}
	\end{bmatrix} .
\end{align}
This grouping of the Killing vectors was introduced in Ref.~\cite{Helset:2022tlf}.
The commutator of covariant derivatives for the ghosts is
\begin{align}
 \left[ \nnewD_{\mu} , \nnewD_{\nu} \right]^{A}_{\ B}= 
 \left[ \mathcal{Y}_{\mu\nu} \right]^A{}_{B} &=  \widetilde R^A{}_{BKL} (D_\mu \phi)^K (D_\nu \phi)^L  + \widetilde\nabla_{B} \widetilde t^{A}_{C}  F^C_{\mu \nu} 
\nonumber \\
&= \widetilde R^A{}_{BKL} (D_\mu \phi)^K (D_\nu \phi)^L   - \f A CB F_{\mu \nu}^C +  \Gamma^A_{LB} t^L_C F^C_{\mu \nu} \,.
\end{align}

The divergent contribution in Eq.~\eqref{eq:divergence} allows us to compute the anomalous dimension of the effective Lagrangian. The remaining computation is purely algebraic. Evaluate $Y_{\mu \nu}$ and $\mathcal{X}$ in terms of the metrics and potential in the Lagrangian, and then take the traces in Eq.~\eqref{eq:divergence}. Note that matrix multiplication and traces are over the combined scalar-gauge space. The ghost contribution is subtracted, since  ghosts are anticommuting. We discuss the application of our results to the SMEFT in the next section.

\section{Standard Model Effective Field Theory}\label{sec:smeft}

Although the construction and main results of this paper apply to a general effective field theory for scalars and gauge fields, it is of particular interest to apply it to the SMEFT. In the SMEFT, the only scalar field is the Higgs doublet, which we write as four real scalars, as in Eq.~\eqref{A.1}, 
\begin{align}
	H = \frac{1}{\sqrt{2}} 
	\begin{pmatrix}
		\phi^2 + i \phi^1 \\
		\phi^4 - i \phi^3
	\end{pmatrix} \,,
\end{align}
and the scalar indices $I,J,\dots$ take values from 1 to 4.  We group all gauge fields of the full gauge group $SU(3)_{c} \otimes SU(2)_{L} \otimes U(1)_{Y}$ into the multiplet
\begin{align}
	A^{B}_{\mu} = 
	\begin{pmatrix}
		G^{\gA}_{\mu} \\
		W^{a}_{\mu} \\
		B_{\mu} 
	\end{pmatrix} .
\end{align}
The corresponding field-strength tensors are $G^{\mathscr{A}}_{\mu \nu}$, $W^a_{\mu \nu}$, and $B_{\mu \nu}$.  Unless otherwise specified, $a$ runs from 1 to 3. At times we will combine the electroweak $SU(2)_{L} \otimes U(1)_{Y}$ gauge groups, and let $a$ run from 1 to 4, where $W^{4}_{\mu\nu} = B_{\mu\nu}$, and denote this explicitly.

The operators in the starting SMEFT Lagrangian are those that can be included as terms in the metrics or potential. All fermions are dropped. The terms in the SMEFT Lagrangian to dimension four are the SM terms
\begin{align}\label{4.3sm}
\mathcal{L} =&  -\frac14 G^{\mathscr{A}}_{\mu \nu} G^{\mathscr{A} \mu \nu}  -\frac14 W^a_{\mu \nu} W^{a \, \mu \nu} - \frac14 B_{\mu \nu} B^{\mu \nu}+ D_\mu H^\dagger D_\mu H - \lambda \left( H^\dagger H - \frac12 v^2\right)^2 \nn
=& -\frac14 G^{\mathscr{A}}_{\mu \nu} G^{\mathscr{A} \mu \nu}  -\frac14 W^a_{\mu \nu} W^{a \, \mu \nu} - \frac14 B_{\mu \nu} B^{\mu \nu} + \frac12 (D_\mu \phi)^I (D^\mu \phi)^I - \frac14 \lambda (\phi^I \phi^I - v^2)^2 \,.
\end{align}
From Eq.~\eqref{4.3sm}, we can read off the potential
\begin{align}\label{4.4}
V(\phi) =& \frac14 \lambda (\phi^I \phi^I - v^2)^2 \,,
\end{align}
and gauge covariant derivative~\cite{Alonso:2015fsp}
\begin{align}\label{4.5}
(D_\mu \phi)^I =\partial_\mu \begin{bmatrix} \phi^1 \\ \phi^2 \\ \phi^3 \\ \phi^4 \end{bmatrix} 
+\frac12 \begin{bmatrix}
0&g _2 W_\mu^3+g_1 B_\mu & -g_2 W_\mu^2&g_2 W_\mu^1\\
-g_2 W_\mu^3-g_1 B_\mu &0& g_2 W_\mu^1&g_2 W_\mu^2\\
g_2 W_\mu^2&-g_2 W_\mu^1&0& g_2 W_\mu^3-g_1 B_\mu \\
-g_2 W_\mu^1&-g_2 W_\mu^2&-g_2 W_\mu^3 +g_1 B_\mu&0\\
\end{bmatrix} \begin{bmatrix} \phi^1 \\ \phi^2 \\ \phi^3 \\ \phi^4 \end{bmatrix} \,.
\end{align}
The Killing vectors $t_a$ can be read off from Eq.~\eqref{4.5}
\begin{align}\label{4.6}
t_1 =& \frac12 g_2 \begin{bmatrix} \phi^4 \\  \phi^3  \\ -\phi_2 \\ - \phi_1 \end{bmatrix} \,, &
t_2 =& \frac12 g_2 \begin{bmatrix} -\phi^3 \\  \phi^4 \\ \phi_1 \\ - \phi_2 \end{bmatrix} \,, &
t_3 =& \frac12 g_2 \begin{bmatrix} \phi^2 \\  -\phi^1 \\ \phi_4 \\ - \phi_3 \end{bmatrix} \,, &
t_4 =& \frac12 g_1 \begin{bmatrix} \phi^2 \\  -\phi^1 \\ -\phi_4 \\ \phi_3 \end{bmatrix} \,,
\end{align}
using $t_{1,2,3}$ for the $SU(2)_L$ generators and $t_4$ for the $U(1)_Y$ generator.

The notation we use for the SMEFT operators is given in App.~\ref{sec:Operators} and follows the notation of Ref.~\cite{Murphy:2020rsh}. We include an additonal left superscript with the operator dimension, since we need operators of dimension $d=2,4,6,8$. The even-parity bosonic SM operators of dimension two and four are given in Tables~\ref{dim:2} and \ref{dim:4}. These are generated by the one-loop formula, Eq.~\eqref{eq:divergence}.

We include all SM couplings other than the Yukawa couplings in our calculation.
%
\begin{table}
\begin{center}
\begin{minipage}[t]{2cm}
\vspace{-0.5cm}
\renewcommand{\arraystretch}{1.5}
\begin{align*}
\begin{array}{c|c}
\multicolumn{2}{c}{\bm{H^2} } \\
\hline
\op{2}{H^2}{} & (H^\dagger H)
\end{array}
\end{align*}
\end{minipage}
\end{center}
\caption{\label{dim:2} Bosonic dimension-two operator in the SM (and the SMEFT).}
\end{table}
%
%
%
\begin{table}
\begin{center}
\begin{minipage}[t]{2.75cm}
\vspace{-0.5cm}
\renewcommand{\arraystretch}{1.5}
\begin{align*}
\begin{array}{c|c}
\multicolumn{2}{c}{\bm{H^4} } \\
\hline
\op{4}{H^4}{} &  (H^\dagger H)^2 \\
\end{array}
\end{align*}
\end{minipage}
\hspace{0.5cm}
\begin{minipage}[t]{5cm}
\vspace{-0.5cm}
\renewcommand{\arraystretch}{1.5}
\begin{align*}
\begin{array}{c|c}
\multicolumn{2}{c}{\bm{H^2 D^2} } \\
\hline
\op{4}{H^2D^2}{} & (D^\mu H^\dagger D_\mu H) \\
\opr{4}{H^2\Box}{} & (H^\dagger D^2 H) + (D^2 H^\dagger H) \\
\end{array}
\end{align*}
\end{minipage}
\hspace{0.5cm}
\begin{minipage}[t]{3.25cm}
\vspace{-0.5cm}
\renewcommand{\arraystretch}{1.5}
\begin{align*}
\begin{array}{c|c}
\multicolumn{2}{c}{\bm{X^2} } \\
\hline
\op{4}{G^2}{} &  G_{\mu \nu}^{\mathscr{A}} G^{\mathscr{A} \mu \nu} \\
\op{4}{W^2}{} &  W_{\mu \nu}^a W^{a \, \mu \nu} \\
\op{4}{B^2}{} &  B_{\mu \nu} B^{\mu \nu}
\end{array}
\end{align*}
\end{minipage}
\end{center}
\caption{\label{dim:4} Bosonic dimension-four operators in the SM (and the SMEFT). The operator $\opr{4}{H^2\Box}{} $ is redundant, and can be eliminated by integration by parts.}
\end{table}
We use the Warsaw basis~\cite{Grzadkowski:2010es} for the dimension-six terms, listed in Table~\ref{dim:6} with our notation and the original notation of Ref.~\cite{Grzadkowski:2010es}.
%
\begin{table}
\begin{center}
\begin{minipage}[t]{5cm}
\vspace{-0.5cm}
\renewcommand{\arraystretch}{1.5}
\begin{align*}
\begin{array}{c|c|c}
\multicolumn{3}{c}{\bm{X^3}} \\
\hline
Q_G     & \op{6}{G^3}{}           & f^{\mathscr{A}\mathscr{B}\mathscr{C}} G_\mu^{\mathscr{A}\nu} G_\nu^{\mathscr{B}\rho} G_\rho^{\mathscr{C}\mu}  \\
Q_W      & \op{6}{W^3}{}            & \epsilon^{abc} W_\mu^{a\,\nu} W_\nu^{b\,\rho} W_\rho^{c \, \mu} \\ 
\end{array}
\end{align*}
\renewcommand{\arraystretch}{1.5}
\begin{align*}
\begin{array}{c|c|c}
\multicolumn{3}{c}{\bm{H^4 D^2} } \\
\hline
Q_{H\Box} & \op{6}{H^4\Box}{} & (H^\dag H)\Box (H^\dag H) \\
Q_{H D} &\op{6}{H^4D^2}{}  & \left(D^\mu H^\dag H\right) \left(H^\dag D_\mu H\right)
\end{array}
\end{align*}
\end{minipage}
\hspace{1cm}
\begin{minipage}[t]{2.5cm}
\vspace{-0.5cm}
\renewcommand{\arraystretch}{1.5}
\begin{align*}
\begin{array}{c|c|c}
\multicolumn{3}{c}{\bm{H^6} } \\
\hline
Q_H    & \op{6}{H^6}{}   & (H^\dag H)^3
\end{array}
\end{align*}
\vspace{-0.5cm}
\renewcommand{\arraystretch}{1.5}
\begin{align*}
\begin{array}{c|c|c}
\multicolumn{3}{c}{\bm{X^2 H^2} } \\
\hline
Q_{H G}  & \op{6}{G^2H^2}{}   & (H^\dag H)\, G^{\mathscr{A}}_{\mu\nu} G^{\mathscr{A}\mu\nu} \\
Q_{H W}  &\op{6}{W^2H^2}{}     & (H^\dag H) \, W^a_{\mu\nu} W^{a\mu\nu} \\
Q_{H B} & \op{6}{B^2H^2}{}      &  (H^\dag H)\, B_{\mu\nu} B^{\mu\nu} \\
Q_{H WB} &  \op{6}{WBH^2}{}    &  (H^\dag \tau^a H)\, W^a_{\mu\nu} B^{\mu\nu} \\
\end{array}
\end{align*}
\end{minipage}
\end{center}
\caption{\label{dim:6} Bosonic even-parity dimension-six operators in the SMEFT. The first column is the notation of Ref.~\cite{Grzadkowski:2010es}, and the second column is the notation used in this paper.}
\end{table}
The $H^6$ SMEFT operator gives a contribution to the scalar potential, the $H^4 D^2$ operators contribute to the scalar metric, and the $X^2 H^2$ operators contribute to the gauge metric, where $X$ is a general field strength. We cannot include the $X^3$ operators in our initial Lagrangian in Eq.~\eqref{eq:Lagr}.
The dimension-eight even-parity bosonic operators are listed in Table~\ref{dim:8}, excluding the $X^4$ operators which we do not need for this paper.
%
\begin{table}
\begin{center}
\begin{minipage}[t]{2.15cm}
\vspace{-0.5cm}
\renewcommand{\arraystretch}{1.5}
\begin{align*}
\begin{array}{c|c}
\multicolumn{2}{c}{\bm{H^8} } \\
\hline
\op{8}{H^8}{} &  (H^\dag H)^4 
\end{array}
\end{align*}
\vspace{-0.5cm}
\renewcommand{\arraystretch}{1.5}
\begin{align*}
\begin{array}{c|c}
\multicolumn{2}{c}{\bm{H^6 D^2} } \\
\hline
\op{8}{H^6 D^2}{(1)}  & (H^{\dag} H)^2 (D_{\mu} H^{\dag} D^{\mu} H) \\
\op{8}{H^6 D^2}{(2)}  & (H^{\dag} H) (H^{\dag} \tau^I H) (D_{\mu} H^{\dag} \tau^I D^{\mu} H)
\end{array}
\end{align*}

\vspace{-0.5cm}
\renewcommand{\arraystretch}{1.5}
\begin{align*}
\begin{array}{c|c}
\multicolumn{2}{c}{\bm{H^4 D^4} } \\
\hline
\op{8}{H^4 D^4}{(1)}  &  (D_{\mu} H^{\dag} D_{\nu} H) (D^{\nu} H^{\dag} D^{\mu} H) \\ 
\op{8}{H^4 D^4}{(2)}  &  (D_{\mu} H^{\dag} D_{\nu} H) (D^{\mu} H^{\dag} D^{\nu} H) \\ 
\op{8}{H^4 D^4}{(3)}  &  (D^{\mu} H^{\dag} D_{\mu} H) (D^{\nu} H^{\dag} D_{\nu} H)
\end{array}
\end{align*}
\vspace{-0.5cm}
\renewcommand{\arraystretch}{1.5}
\begin{align*}
\begin{array}{c|c}
\multicolumn{2}{c}{\bm{X^3 H^2} } \\
\hline
\op{8}{G^3H^2}{(1)}  &  f^{\mathscr{A}\mathscr{B}\mathscr{C}} (H^\dag H) G_{\mu}^{\mathscr{A}\nu} G_{\nu}^{\mathscr{B}\rho} G_{\rho}^{\mathscr{C}\mu} \\
\op{8}{W^3H^2}{(1)}  &  \epsilon^{abc} (H^\dag H) W_{\mu}^{a\,\nu} W_{\nu}^{b\,\rho} W_{\rho}^{c\,\mu} \\
\op{8}{W^2BH^2}{(1)}  &  \epsilon^{abc} (H^\dag \tau^a H) B_{\mu}^{\;\;\,\nu} W_{\nu}^{b\,\rho} W_{\rho}^{c\,\mu} \\
\end{array}
\end{align*}
\end{minipage}
\hspace{0.5cm}
\begin{minipage}[t]{4cm}
\vspace{-0.5cm}
\renewcommand{\arraystretch}{1.5}
\begin{align*}
\begin{array}{c|c}
\multicolumn{2}{c}{\bm{X^2 H^4} } \\
\hline
\op{8}{G^2H^4}{(1)}  & (H^\dag H)^2 G^{\mathscr{A}}_{\mu\nu} G^{\mathscr{A}\mu\nu} \\
\op{8}{W^2H^4}{(1)}  & (H^\dag H)^2 W^a_{\mu\nu} W^{a\,\mu\nu} \\
\op{8}{W^2H^4}{(3)}  & (H^\dag \tau^a H) (H^\dag \tau^b H) W^a_{\mu\nu} W^{b\,\mu\nu} \\
\op{8}{WBH^4}{(1)}  &  (H^\dag H) (H^\dag \tau^a H) W^a_{\mu\nu} B^{\mu\nu} \\
\op{8}{B^2H^4}{(1)}  &  (H^\dag H)^2 B_{\mu\nu} B^{\mu\nu} \\
\end{array}
\end{align*}
\vspace{-0.5cm}
\renewcommand{\arraystretch}{1.5}
\begin{align*}
\begin{array}{c|c}
\multicolumn{2}{c}{\bm{X  H^4 D^2} } \\
\hline
\op{8}{WH^4D^2}{(1)}  & i (H^{\dag} H) (D^{\mu} H^{\dag} \tau^a D^{\nu} H) W_{\mu\nu}^a \\
\op{8}{WH^4D^2}{(3)}  & i \epsilon^{abc} (H^{\dag} \tau^a H) (D^{\mu} H^{\dag} \tau^b D^{\nu} H) W_{\mu\nu}^c \\
\op{8}{BH^4D^2}{(1)}  & i (H^{\dag} H) (D^{\mu} H^{\dag} D^{\nu} H) B_{\mu\nu} \\
\end{array}
\end{align*}
\end{minipage}
\vspace{-0.5cm}
\renewcommand{\arraystretch}{1.5}
\begin{align*}
\begin{array}{c|c}
\multicolumn{2}{c}{\bm{X^2 H^2 D^2} } \\
\hline
\op{8}{G^2H^2D^2}{(1)}  &  (D^{\mu} H^{\dag} D^{\nu} H) G_{\mu\rho}^{\mathscr{A}} G_{\nu}^{\mathscr{A} \rho} \\
\op{8}{G^2H^2D^2}{(2)}  &  (D^{\mu} H^{\dag} D_{\mu} H) G_{\nu\rho}^{\mathscr{A}} G^{\mathscr{A} \nu\rho} \\
\op{8}{W^2H^2D^2}{(1)}  &  (D^{\mu} H^{\dag} D^{\nu} H) W_{\mu\rho}^a W_{\nu}^{a\, \rho} \\
\op{8}{W^2H^2D^2}{(2)}  &  (D^{\mu} H^{\dag} D_{\mu} H) W_{\nu\rho}^a W^{a\, \nu\rho} \\
\op{8}{W^2H^2D^2}{(4)}  &  i \epsilon^{abc} (D^{\mu} H^{\dag} \tau^a D^{\nu} H) W_{\mu\rho}^b W_{\nu}^{c\, \rho} \\
\op{8}{WBH^2D^2}{(1)}  &  (D^{\mu} H^{\dag} \tau^a D_{\mu} H) B_{\nu\rho} W^{a\, \nu\rho} \\
\op{8}{WBH^2D^2}{(3)}  &  i (D^{\mu} H^{\dag} \tau^a D^{\nu} H) (B_{\mu\rho} W_{\nu}^{a\, \rho} - B_{\nu\rho} W_{\mu}^{a\,\rho}) \\
\op{8}{WBH^2D^2}{(4)}  &  (D^{\mu} H^{\dag} \tau^a D^{\nu} H) (B_{\mu\rho} W_{\nu}^{a\, \rho} + B_{\nu\rho} W_{\mu}^{a\,\rho}) \\
\op{8}{B^2H^2D^2}{(1)}  &  (D^{\mu} H^{\dag} D^{\nu} H) B_{\mu\rho} B_{\nu}^{\,\,\,\rho} \\
\op{8}{B^2H^2D^2}{(2)}  &  (D^{\mu} H^{\dag} D_{\mu} H) B_{\nu\rho} B^{\nu\rho} \\
\end{array}
\end{align*}

\hspace{-1.5cm}
%
\end{center}
\caption{\label{dim:8} Bosonic dimension-eight operators in the SMEFT.  The $XH^4D^2$ operators have a factor of $i$ relative to Ref.~\cite{Murphy:2020rsh} to make them hermitian. There are also $X^4$ operators which have not been listed.}
\end{table}

The $H^8$ operator contributes to the potential, the $H^6 D^2$ operators to the scalar metric, and $X^2 H^4$ operators to the gauge metric.

The total potential to dimension eight is
\begin{align}
V =&  \frac14 \lambda (\phi^I \phi^I - v^2)^2 -\frac18\ \coef{6}{H^6}{} (\phi^I \phi^I)^3-\frac1{16}\ \coef{8}{H^8}{} (\phi^I \phi^I)^4 \,,
\end{align}
and the total scalar metric to dimension eight is
\begin{align}\label{5.10}
h_{IJ} = & \delta_{IJ} \left[ 1+ \frac 14 \left( \coef{8}{H^6 D^2}{(1)} - \coef{8}{H^6 D^2}{(2)}\right) (\phi^K \phi^K)^2  \right] +\left( -2\ \ \coef{6}{H^4\Box}{}  \right) \phi^I \phi^J \nn
& + \frac12 \left[ \coef{6}{H^4D^2}{} + \coef{8}{H^6 D^2}{(2)} (\phi^K \phi^K)  \right]  \mathcal{H}_{IJ}(\phi) \,,
\end{align}
where
\begin{align}
\mathcal{H}_{IJ}(\phi) =& \phi_I \phi_J +  \begin{bmatrix}
\phi_2^2 & -\phi_1 \phi_2 & - \phi_2 \phi_4 & \phi_2 \phi_3  \\
-\phi_1 \phi_2 & \phi_1^2 & \phi_1 \phi_4 & - \phi_1 \phi_3 \\
-\phi_2 \phi_4 & \phi_1 \phi_4 & \phi_4^2 & - \phi_3 \phi_4 \\
\phi_2 \phi_3 & - \phi_1 \phi_3 & -\phi_3 \phi_4 & \phi_3^2
\end{bmatrix} \,.
\end{align}
The matrix $\mathcal{H}$ is
\begin{align}
\mathcal{H}_{IJ}(\phi) =& \frac12 \sum_{a=1}^4 \left[\Upsilon_a \right]_{IJ} x_a(\phi) \,, & x_a(\phi) =& \left[\Upsilon_a \right]_{KL} \phi^K \phi^L \,,
\end{align}
where the matrices $\Upsilon$ were defined in Ref.~\cite{Helset:2020yio}, and are discussed in App.~\ref{sec:higgs}.

The total gauge metric is
\begin{align}
g_{AB} =& \left[ \begin{array}{ccc} 
[g_{GG}]_{\mathscr{A} \mathscr{B}}  & 0 & 0 \\
		0 & [g_{WW}]_{ab} & [g_{WB}]_a  \\ 
		0 & [g_{BW}]_b & g_{BB}  
\end{array} \right] \,,
\end{align}
where $A$ and $B$ run over the $SU(3)$, $SU(2)$, and $U(1)$ gauge groups. The submatrices are
\begin{align}
	\label{eq:gMetricSMEFT}
g_{GG} =& \left[ 1 - 2\ \ \coef{6}{G^2H^2}{} (\phi^I \phi^I) -   \coef{8}{G^2H^4}{} (\phi^I \phi^I)^2 \right] \mathbbm{1}_{8 \times 8} \, , \nn
[g_{WW}]_{ab} =&   \left[ 1 - 2\ \ \coef{6}{W^2H^2}{} (\phi^I \phi^I) -   \coef{8}{W^2H^4}{(1)} (\phi^I \phi^I)^2 \right] \mathbbm{1}_{3 \times 3}  -4 \ \ \coef{8}{W^2H^4}{(2)} x_a(\phi) x_b(\phi) \, , \nn
[g_{WB}]_a =& [g_{BW}]_a =   \left( 2\ \ \coef{6}{WBH^2}{}  + \coef{8}{WBH^4}{}   \right) x_a \,, \nn
g_{BB} =&  \left[ 1 - 2\ \ \coef{6}{B^2H^2}{} (\phi^I \phi^I) -   \coef{8}{B^2H^4}{} (\phi^I \phi^I)^2 \right]   \,.
\end{align}
Note that $B$ in Eq.~\eqref{eq:gMetricSMEFT} denotes the $B_{\mu\nu}$ field strength for the $U(1)_Y$ gauge group, and not a gauge index.
From Eq.~\eqref{5.10},
\begin{align}
	R_{I J K L} =& -2\ \ \coef{6}{H^4\Box}{} (\delta_{IK} \delta_{JL} - \delta_{IL} \delta_{JK} ) -  \frac12  \coef{6}{H^4D^2}{} \sum_{a=1}^{4} \left(  [\Upsilon_a]_{IK} [\Upsilon_a]_{JL} - [\Upsilon_a]_{IL} [\Upsilon_a]_{JK} \right) \nn
& -  \left( 4 \left( \coef{6}{H^4\Box}{} \right)^2 +\coef{8}{H^6 D^2}{(1)} - \coef{8}{H^6 D^2}{(2)}\right)  (\phi^R \phi^R) (\delta_{IK} \delta_{JL} - \delta_{IL} \delta_{JK} ) \nn
& + \left( \coef{8}{H^6 D^2}{(1)} - \coef{8}{H^6 D^2}{(2)}\right) \left( \delta_{JK} \phi_I \phi_L + \delta_{I L} \phi_J \phi_K - \delta_{JL} \phi_I \phi_K - \delta_{I K} \phi_J \phi_L \right) \nn
&+ \text{dimension-eight $\Upsilon$ terms}\,,
\label{3.18a}
\end{align}
where the $\Upsilon$ terms are complicated, and have not been shown explicitly.

\section{Renormalization Group Equations}\label{sec:rge}

The RGEs can be computed using Eq.~\eqref{eq:divergence}. The starting Lagrangian to dimension eight is given by expanding the potential and the scalar and gauge metrics to order $1/M^4$. The coefficients included in the Lagrangian are those that multiply operators given by expanding the potential and the metrics in Eq.~\eqref{eq:Lagr}. As we have seen, the SM couplings $m_H^2$, $\lambda$, $g_1$, $g_2$, and $g_3$ can all be included, so the only SM couplings which are dropped are the Yukawa couplings.
The dimension-six SMEFT coefficients included in the form in Eq.~\eqref{eq:Lagr} are
\begin{align}\label{5.1}
& \coef{6}{H^6}{},\ \coef{6}{H^4\Box}{},\ \coef{6}{H^4D^2}{},\ \coef{6}{G^2H^2}{},\ \coef{6}{W^2H^2}{},\ \coef{6}{B^2H^2}{},\ {\coef{6}{WBH^2}{}},
\end{align}
and the dimension-eight coefficients are
\begin{align}
\coef{8}{H^8}{},\  \coef{8}{H^6D^2}{(1)},\ \coef{8}{H^6D^2}{(2)},\ \coef{8}{G^2H^4}{(1)}, \ \coef{8}{W^2H^4}{(1)},\ \coef{8}{W^2H^4}{(3)},\ \coef{8}{B^2H^4}{(1)},\ \coef{8}{WBH^4}{(1)}\,.
\label{5.2}
\end{align}
While the starting Lagrangian has the form in Eq.~\eqref{eq:Lagr}, the counterterms generated are not limited to that form. Thus we can compute all  terms in the entire RGE to dimension eight which depend on the coefficients listed above.\footnote{Warning: As an example, the RGEs for $H^4D^2$ operators in App.~\ref{sec:RGEresults} do not include $H^4D^4$ terms, since these are not in the listed coefficients. This does not mean that such terms vanish. They are present, e.g., from wavefunction renormalization contributions, and were computed in Ref.~\cite{DasBakshi:2022mwk}. Similarly for all other contributions which depend on coefficients not listed in Eqs.~\eqref{5.1} and~\eqref{5.2}.}
 As a non-trivial check of the method, we reproduce the known SM RGEs, and the dimension-six RGEs computed in Refs.~\cite{Jenkins:2013zja,Jenkins:2013wua,Alonso:2013hga} which depend on the above coefficients. The dimension-four and dimension-six RGEs are not included in the results in App.~\ref{sec:RGEresults}, and must be added to the equations given there.

The operators generated by the renormalization counterterms are not in the canonical operator basis. The most lengthy part of the computation is converting the redundant operators to the canonical basis using integration-by-parts and field-redefinition identities. These relations are tabulated in App.~\ref{sec:Operators}. Eliminating operators by field redefinitions is equivalent to using the equations of motion to first order, but not to higher orders~(see, e.g., Ref.~\cite{Manohar:1997qy} for an HQET example). In our computation, the redundant operators have a one-loop coefficient, so second-order quantities are effectively two-loop order, and can be dropped. Thus one can use the classical equations of motion instead of field redefinitions to get the results in App.~\ref{sec:Operators}. This simplification cannot be used for a two-loop computation.

The RGEs are given in App.~\ref{sec:RGEresults} for the dimension-eight corrections to the SM couplings $m_H^2$ and $\lambda$, and to the dimension-six SMEFT couplings $\coef{6}{H^6}{}$, $\coef{6}{H^4\Box}{}$, $\coef{6}{H^4D^2}{}$, $\coef{6}{G^2H^2}{}$,  $\coef{6}{W^2H^2}{}$, $\coef{6}{B^2H^2}{}$,  and $\coef{6}{WBH^2}{}$. We also obtain the dimension-eight RGEs for the $H^8$, $H^6D^2$, $H^4D^4$, $X^2H^4$, $X^3H^2$, $X^2H^2D^2$, and $XH^4D^2$ operators. All other dimension-eight SMEFT coefficients have no terms in their RGE that depend on the coefficients in Eqs.~\eqref{5.1}~and~\eqref{5.2} and the SM bosonic couplings. Following the convention of Refs.~\cite{Jenkins:2013zja,Jenkins:2013wua,Alonso:2013hga}, we list $16\pi^2 \mu \tfrac{\rd}{\rd \mu}$ for the RGEs to avoid a factor of $\tfrac{1}{16 \pi^2}$ in each equation.

Parts of the dimension-eight RGEs have been computed previously in Refs.~\cite{Chala:2021pll,DasBakshi:2022mwk}. These results  include fermionic terms which we have not computed, and we have some bosonic contributions which they have not computed. Ref.~\cite{DasBakshi:2022mwk} computes the RGEs for the dimension-eight operators proportional to dimension-eight coefficients. We agree with their results for all the terms both calculations have in common. In comparing, it is necessary to take into account the opposite sign convention for the coupling constant in the gauge covariant derivative, and the difference in gauge $\beta$-functions because we do not include fermion loops. 

The one-loop correction to the SM kinetic terms gives the field anomalous dimensions  (wavefunction renormalization factors) listed in Eq.~\eqref{C.3}. These depend on the gauge-fixing term, and are computed using our choice in Eq.~\eqref{4.21}. There are dimension-six corrections to the wavefunction renormalization. Note that the dimension-four contribution to $\gamma_H$, $\gamma_H= -g_1^2-3g_2^2$, is twice the value in Feynman gauge.

As in previous calculations~\cite{Jenkins:2013zja}, there are some unusually large coefficients in the RGEs, with several coefficients larger than 100. The biggest coefficient is in the RGE
\begin{align}
\mu \frac{\rd}{\rd \mu} \coef{8}{H^8}{} =& - \frac{1568}{16\pi^2} \lambda^2 \left( \coef{6}{H^4\Box}{} \right)^2  + \ldots 
\approx - 0.17  \left( \coef{6}{H^4\Box}{} \right)^2  + \ldots 
\end{align}
using the known value of $\lambda$ in the SM.

\subsection{NDA}

There are some consistency checks on our result using the NDA~\cite{Manohar:1983md} rules given in Refs.~\cite{Jenkins:2013sda,Gavela:2016bzc}. Every operator in the Lagrangian has mass dimension $d$, and an NDA weight $u=F-2$, where $F$ is the number of fields in the operator.\footnote{$u$ is  $2w$, where $w$ was defined in Ref.~\cite{Jenkins:2013sda}. This avoids half-integer values.} One can write a Lagrangian term as
\begin{align}
L = & C O = \widehat C \widehat O \,,
\end{align}
where
\begin{align}
\widehat O =& \frac{(4\pi)^u}{M^{d-4}} O \,, & \widehat C =&  \frac{M^{d-4}}{(4\pi)^u} C \,,
\end{align}
and are dimensionless, and $M$ is the SMEFT power counting scale. For example,
\begin{align}
L = & \coef{8}{H^6D^2}{(1)} (H^\dagger H)^2(D_\mu H^\dagger D^\mu H)  +  \coef{8}{H^8}{} H^8 \nn
 = & \left[ \frac{M^{4}}{(4\pi)^4}   \coef{8}{H^6D^2}{(1)} \right] \left[ \frac{(4\pi)^4}{M^{4}}  (H^\dagger H)^2(D_\mu H^\dagger D^\mu H)  \right]  +  \left[ \frac{M^{4}}{(4\pi)^6}    \coef{8}{H^8}{} \right]  \left[ \frac{(4\pi)^6}{M^{4}} H^8 \right]
\end{align}
so that
\begin{align}
\hcoef{8}{H^6D^2}{(1)}=& \frac{M^{4}}{(4\pi)^4}   \coef{8}{H^6D^2}{(1)} \,, &
 \hcoef{8}{H^8}{}  =& \frac{M^{4}}{(4\pi)^6}    \coef{8}{H^8}{} \,.
\end{align}
Then if an operator $O$ is generated by an $L$-loop graph with insertions of operators $O_i$, we have the power counting rules~\cite{Manohar:1983md,Jenkins:2013sda,Gavela:2016bzc}
\begin{align}\label{5.7}
\mu\frac{\rd}{\rd \mu} {\hat C} \sim \prod_i \hat C_i \,,
\end{align}
where the overall coefficient is order one, and
\begin{align}\label{5.8}
d-4 =& \sum_i (d_i-4)\,, &
u +2 L =& \sum_i u_i \,.
\end{align}
A strongly coupled theory is one where $\hat C_i$ are order unity. It is inconsistent to have theories with $\hat C_i$ much larger than unity. In a weakly coupled theory, $\hat C_i$ can be much smaller than unity, and Eq.~\eqref{5.7} implies that loop corrections are small.

The $H^8$ operator has $d=8$, $u=6$, the $H^6 D^2$ and $X^2 H^4$ operators have $d=8$, $u=4$, the $H^6$ operator has $d=6$, $u=4$, the $H^4D^2$ and $X^2 H^2$ operators have $d=6$ and $u=2$, the $H^4$ operator which multiples $\lambda$ has $d=4$, $u=2$, and the gauge couplings multiply a $d=4$, $u=1$ operator. Thus from Eq.~\eqref{5.8}, the one-loop RGE for $H^8$ can only have terms of the form 
\begin{align}
	\dcoef{8}{H^8}{} \sim & \; \left(\lambda, g^2\right) \times \left\{ \coef{8}{H^8}{} \right\} \; + \; \left(\lambda , \, g^2\right)^2 \times \left\{ \coef{8}{H^6 D^2}{} , \,  \coef{8}{X^2 H^4 }{} \right\}  \; + \; \left( \coef{6}{H^6}{} \right)^2 
	\nn &
	\; + \;  \left(\lambda,\, g^2\right) \times  \coef{6}{H^6}{}  \times \left\{ \coef{6}{X^2 H^2}{} , \; \coef{6}{H^4D^2}{} \right\} \; + \; 
\left(\lambda,\,  g^2\right)^2 \times \left\{ \coef{6}{X^2 H^2}{} , \; \coef{6}{H^4D^2}{} \right\}^2  \,,
\end{align}
which is the structure of the terms found in Eq.~\eqref{C.21}. One can similarly check that the other RGEs satisfy Eq.~\eqref{5.8}.

\section{Geometric Zeros}\label{sec:Zeros}

Much interest has centered around some surprising vanishing terms in the RGEs for the SMEFT,  first found in Ref.~\cite{Alonso:2014rga}.  For example, using the NDA described above for the dimension-six coefficient $\coef{6}{H^4 D^2}{}$, the RGE can depend on
\begin{align}
	\dcoef{6}{H^4D^2}{} \sim \left(\lambda,\; g^2\right) \times \left\{ \coef{6}{H^4 D^2}{},\; \coef{6}{X^2 H^2}{} \right\} \,.
\end{align}
The RGE for the dimension-eight coefficient $\coef{8}{H^6 D^2}{}$  can depend on
\begin{align}
	\dcoef{8}{H^6D^2}{} \sim & \; \left(\lambda,\; g^2\right) \times \left\{ \coef{8}{H^6 D^2}{},\; \coef{8}{X^2 H^4}{} \right\} + \left(\lambda ,\; g^2\right) \times \left\{ \coef{6}{H^4 D^2}{},\; \coef{6}{X^2 H^2}{} \right\}^2 
	\nn &
	+ \coef{6}{H^6}{} \times \left\{ \coef{6}{H^4 D^2}{},\; \coef{6}{X^2 H^2}{} \right\} \,.
\end{align}
However, some of these entries vanish non-trivially---$(g^2) \times \coef{6}{X^2 H^2}{}$ in the first example, and $(g^2) \times \coef{8}{X^2 H^4}{}$ in the second example---when all the diagrams are added up.
The zeros at dimension six were explained using on-shell methods \cite{Cheung:2015aba}. Similar vanishing entries in the RGEs have also been encountered at two loops \cite{Bern:2020ikv} and when including gravity \cite{Baratella:2021guc}.

We will explore this using the geometric approach to the RGE. 
Let us focus on the terms of the schematic form $(t\cdot t) (\nabla^2 g) \;  (\partial \phi)^2$, which would contribute to the RGE of the form $\dcoef{d}{H^{d-2} D^2}{} \sim (g^2) \times \coef{d}{X^2 H^{d-4}}{}$. The gauge variation gives
\begin{align}
	\frac{1}{2} {\rm Tr}[ X^2 ]_{\rm gauge}\Biggr|_{(t\cdot t) (\nabla^2 g) \;  (\partial \phi)^2} & =  
	\left[ h_{KL} t^{KA}t^{LB} \eta^{\mu \nu} \right]
	\nn & \times
	\left[ - \frac{1}{2} \nabla_{I} \nabla_{J} g_{AB} \eta_{\mu\nu}
  (D_\alpha \phi)^I   (D_\alpha \phi)^J         
	 + ( \nabla_I \nabla_J g_{AB})  (D_\mu \phi)^I (D_\nu \phi)^J    \right]
	 \nn &=
	 -\left[ t^{A}_{K}t^{KB}  \right]
	\left[ \nabla_{I} \nabla_{J} g_{AB} (D_\alpha \phi)^I   (D^\alpha \phi)^J         \right] \,.
\end{align}
The ghost terms give
\begin{align}
	\frac{1}{2} {\rm Tr}[ X^2 ]_{\rm ghost}\Biggr|_{(t\cdot t) (\nabla^2 g) \;  (\partial \phi)^2}  &=  
	(-2)\left[g^{AC} \frac12 \nabla_I \nabla_J g_{CB}  (D_{\mu} \phi)^{I} (D^{\mu} \phi)^{J} 
	\right] \left[
	- t^{KB} t_{KA} \right]
	 \nn &=
	 \left[ t^{A}_{K}t^{KB}  \right]
	\left[ \nabla_{I} \nabla_{J} g_{AB} (D_\alpha \phi)^I   (D^\alpha \phi)^J         \right] \,.
\end{align}
The ghost terms enter the one-loop calculation with a minus sign, because they are anticommuting, which has been included. This gives the cancellation
\begin{align}
	\frac{1}{2} {\rm Tr}[ X^2 ]_{\rm gauge}\Biggr|_{(t\cdot t) (\nabla^2 g) \;  (\partial \phi)^2}  +  
	\frac{1}{2} {\rm Tr}[ X^2 ]_{\rm ghost}\Biggr|_{(t\cdot t) (\nabla^2 g) \;  (\partial \phi)^2}  &=  0 .
\end{align}
This is consistent with the vanishing of these entries in the RGE that was found at dimension six and dimension eight. In the geometric approach it is clear that the same cancellation happens at all mass dimensions.

\section{Conclusion}\label{sec:Conclusion}

The structure of scattering amplitudes can be elucidated using the geometry of field space. The geometric approach allows for more compact,  field-redefinition-independent expressions for the scattering amplitudes. The geometric approach also simplifies the functional integral calculation of loop corrections.

We have applied the geometry of field space for an effective field theory of scalars and gauge bosons and calculated the one-loop counterterms. The resulting expression depends on geometric quantities, demonstrating that the geometric approach continues to be valuable at loop level.

As an important application of our results, we have calculated the RGEs for the SMEFT for even-parity bosonic operators with mass dimension eight. The results are listed in App.~\ref{sec:RGEresults}.  Some of these terms were previously calculated in the literature~\cite{Chala:2021pll,DasBakshi:2022mwk}, and we agree with previous results. We also have many terms which are new.

Our calculation focused on combining the scalars and gauge bosons into a unified framework. Fermions were not included. However, there should be a geometric formulation for fermions, since field redefinitions involving fermions also leave physical results unchanged. Including fermions, as well as higher-derivative operators, is necessary to capture the full RGEs for the SMEFT at dimension eight. This will be explored in future work.

\subsection*{Acknowledgments}

We thank Clifford Cheung, Xiaochuan Lu, Julio Parra-Martinez, Julie Pag\`es, Chia-Hsien Shen, and Michael Trott for helpful discussions.
We also thank M.~Chala, G.~Guedes, M.~Ramos, and J.~Santiago for confirming some of our RGE results.
This work is supported in part by the U.S.\ Department of Energy (DOE) under award numbers~DE-SC0009919 and DE-SC0011632 and by the Walter Burke Institute for Theoretical Physics.

\appendix

\section{$SU(2)_L \times SU(2)_R$ generators}\label{sec:higgs}

The Higgs doublet $H_i$ and $\widetilde H_i = \epsilon_{ij} H^{\dagger j}$, with $\epsilon_{12}=1$, $\epsilon_{ij}=-\epsilon_{ji}$, both transform as $SU(2)$ doublets, and can be combined into a $2 \times 2$ matrix which can be written as a linear combination of $\mathbbm{1}$ and the Pauli matrices,
\begin{align}\label{A.1}
	\Sigma = \begin{pmatrix} \widetilde H & H \end{pmatrix} = \begin{pmatrix} H^{0*} & H^+ \\ - H^{+*} & H^0  \end{pmatrix}
	= \phi_4 + i \bm{\tau \cdot \phi}\,.
\end{align}
The group $SU(2)_L \times SU(2)_R$ acts on $\Sigma$ by $\Sigma \to g_L \Sigma g_R^\dagger$, giving the generators in the four-dimensional space of $\phi^I$,
\begin{align}
T_L^1 &= \frac{i}{2} \begin{bmatrix} 0 & 0 & 0 & -1 \\ 0 & 0 & -1 & 0 \\ 0 & 1 & 0 & 0 \\ 1 & 0 & 0 & 0\end{bmatrix} \,, &
T_L^2 &= \frac{i}{2} \begin{bmatrix} 0 & 0 & 1 & 0 \\ 0 & 0 & 0 & -1 \\ -1 & 0 & 0 & 0 \\ 0 & 1 & 0 & 0\end{bmatrix}  \,, &
T_L^3 &= \frac{i}{2} \begin{bmatrix} 0 & -1 & 0 & 0 \\ 1 & 0 & 0 & 0 \\ 0 & 0 & 0 & -1 \\ 0 & 0 & 1 & 0\end{bmatrix} \,, \nn
T_R^1 &= \frac{i}{2} \begin{bmatrix} 0 & 0 & 0 & 1 \\ 0 & 0  & -1 & 0  \\ 0 & 1 & 0 & 0 \\ -1 & 0 & 0 & 0 \end{bmatrix} \,, &
T_R^2 &= \frac{i}{2} \begin{bmatrix} 0 & 0 & 1 & 0 \\ 0 & 0  & 0 & 1  \\  -1 & 0 & 0 & 0 \\ 0 & -1 & 0 & 0 \end{bmatrix} \,, &
T_R^3 &= \frac{i}{2} \begin{bmatrix} 0 & -1 & 0 & 0 \\ 1 & 0  & 0 & 0  \\  0 & 0 & 0 & 1 \\ 0 & 0 & -1 & 0 \end{bmatrix} \,.
\end{align}
Defining a fourth Pauli matrix $\tau^4 = -i \mathbbm{1}_{2\times 2}$,
\begin{align}
[T^a_L]^I{}_J =& \frac14 \tr \left( \tau^{I \dagger} \tau^a \tau^J \right) \,, &
[T^a_R]^I{}_J =& -\frac14 \tr \left( \tau^{I \dagger}  \tau^J \tau^a \right) \,.
\end{align}
The Killing vectors in Eq.~\eqref{4.3sm} are
\begin{align}
t^I_a =& i g_2 \left[ T_L^a \right]^I{}_J \ \phi^J , \quad a=1,2,3, &
t^I_4 =& i g_1 \left[ T_R^3 \right]^I{}_J\ \phi^J \,.
\end{align}
It is convenient to define the matrices
\begin{align}
\Upsilon_a = & -4\ T^a_L T^3_R, \qquad a=1,2,3, & 
\Upsilon_4 = & -4 \ T^3_R T^3_R, 
\end{align}
\begin{align}
	\Upsilon_1 = \begin{bmatrix}
		0 & 0 & 1 & 0 \\
		0 & 0 & 0 & -1 \\
		1 & 0 & 0 & 0 \\
		0 & -1 & 0 & 0
	\end{bmatrix}, \quad
	\Upsilon_2  = \begin{bmatrix}
		0 & 0 & 0 & 1 \\
		0 & 0 & 1 & 0 \\
		0 & 1 & 0 & 0 \\
		1 & 0 & 0 & 0
	\end{bmatrix}, \quad
	\Upsilon_3 = \begin{bmatrix}
		-1 & 0 & 0 & 0 \\
		0 & -1 & 0 & 0 \\
		0 & 0 & 1 & 0 \\
		0 & 0 & 0 & 1
	\end{bmatrix},  \quad
	\Upsilon_4 = - \mathbb{I}_{4\times 4} .
\end{align}
These are identical to the matrices $\Gamma_a$ in Ref.~\cite{Helset:2020yio}. We choose to denote them by $\Upsilon$ to avoid confusion with the Christoffel symbol.

\section{Notation and Operator Relations}\label{sec:Operators}

In this paper, we need the bosonic even-parity SMEFT operators up to dimension eight. The renormalization counterterms generate operators of dimension $d=2,4,6,8$ multiplied by $m_H^{8-d}$. In addition to the operators present in the SMEFT Lagrangian in a canonical basis, redundant operators are generated which can be eliminated by integration by parts and by field redefinitions.   At dimension six, we use the Warsaw basis~\cite{Grzadkowski:2010es}, and at dimension eight, the basis in Ref.~\cite{Murphy:2020rsh}.  To avoid confusion, we have used the notation $\op{d}{F}{(n)}$ for operators in the canonical basis. Here $d$ is the operator dimension, $F$ is the field content, and $n$ is used if there are multiple operators with the same dimension and field content. The subscript $F$ is written as a product of powers of $G$, $W$, $B$, $H$, $D$, and $\Box$ for the field-strength tensors, Higgs field, covariant derivatives, and covariant Laplacian. $d$ is redundant, since it can be computed from $F$, but we have included it for clarity. A generic gauge field strength will be denoted by $X$. Other than the left superscript $d$, the notation is that of Ref.~\cite{Murphy:2020rsh}. The coefficient of $\op{d}{F}{(n)}$ in the Lagrangian is denoted by $\coef{d}{F}{(n)}$. 

Redundant operators, which are eliminated by integration by parts and field redefinitions, follow the same notation $\opr{d}{F}{(n)}$ but are denoted by $R$ instead of $Q$.  Redundant operators do not appear in the Lagrangian, so we do not need a notation for their Lagrangian coefficients. Redundant operators are generated at one-loop order. As a result, second-order terms in field redefinitions are of two-loop order, and field redefinitions are equivalent to using the equations of motion accurate up to dimension eight.\footnote{If redundant operators are generated at tree level, e.g.,\ when integrating out a heavy particle, one needs to keep the nonlinear terms in the field redefinitions~\cite{Manohar:1997qy}.} To save space, we have given relations which can be used to reduce each redundant operator to a linear combination of canonical operators, rather than write each redundant operator in terms of canonical operators. For example, $\opr{6}{H^4D^2}{(1)}$ can be written in terms of canonical operators by first using Eq.~\eqref{B.10} followed by Eq.~\eqref{B.8}.

To simplify the notation, it is convenient to define
\begin{align}
 j_{\mu}  = &\  i  (H^\dagger D_\mu H - D_\mu H^\dagger H ) =  i H^\dagger \overleftrightarrow{D}_\mu H \,,&
r_\mu  = &\  \partial_\mu (H^\dagger H)\,, \nn
j^a_{\mu}  = &\  i(H^\dagger \tau^a D_\mu H - D_\mu H^\dagger \tau^a H ) =  i H^\dagger  \overleftrightarrow{D}^a_\mu H\,, &
r^a_\mu  = &\ D_\mu (H^\dagger \tau^a H)\,.
\label{5b}
\end{align}
The commutation relations
\begin{align}
\left[ D_\mu , D_\nu \right] H = &\  \frac{i}{2} g_1  B_{\mu \nu} H + \frac{i}2 g_2 W_{\mu \nu}^a \tau^a H \,, \nn
\left[ D_\mu , D_\nu \right] W^a_{\alpha \beta} =  &\ -g_2 \epsilon^{abc} W_{\mu \nu}^b W_{\alpha \beta}^c\,, \nn
\left[ D_\mu , D_\nu \right] B_{\alpha \beta} = &\  0\,,
\end{align}
can be used to change the order of covariant derivatives. Since $H$ is the only field with a $U(1)_{Y}$ charge in our analysis, the results for a scalar with general hypercharge $\hyp_H$ are given by the replacement $g_1 \to 2 g_1 \hyp_H$.

\subsection{Dimension 0}

At dimension zero, the only operator is $\mathbbm{1}$ with coefficient the negative of the cosmological constant,
\begin{align}
L = - \Lambda + \ldots 
\end{align}

\subsection{Dimension 2}

At dimension two, the only operator is $\op{2}{H^2}{} = (H^\dagger H)$ listed in Table~\ref{dim:2} with coefficient
\begin{align}
L =  \coef{2}{H^2}{}   \op{2}{H^2}{}  =  \frac12 m_H^2 (H^\dagger H)
\end{align}
where $m_H$ is the Higgs mass in the broken phase.

\subsection{Dimension 4}

At dimension four, the canonical bosonic operators are $\op{4}{H^4}{} =  (H^\dagger H)^2$,  $\op{4}{H^2D^2}{} = (D^\mu H^\dagger D_\mu H)$, 
$\op{4}{G^2}{}  = G_{\mu \nu}^{\mathscr{A}} G^{\mathscr{A} \mu \nu} $, $\op{4}{W^2}{}  = W_{\mu \nu}^a W^{a \, \mu \nu} $, and $\op{4}{B^2}{}  = B_{\mu \nu} B^{\mu \nu}$
with
\begin{align}
L =& \sum_{X=G,W,B}  \coef{4}{X^2}{}  \op{4}{X^2}{}   + \coef{4}{H^2D^2}{} \op{4}{H^2D^2}{} +\coef{4}{H^4}{} \op{4}{H^4}{}  \,.
\end{align}
The Higgs self-coupling is $\lambda = - \coef{4}{H^4}{} $.
The fields are rescaled so that the kinetic terms have canonical normalization, $\coef{4}{X^2}{}=-1/4$, $\coef{4}{H^2D^2}{}=1$,  so the anomalous dimensions of $\coef{4}{X^2}{}$ and $\coef{4}{H^2D^2}{}$ give the field anomalous dimensions
\begin{align}
\gamma_X =& 2 \,  \frac{\rd}{\rd \ln \mu}\coef{4}{X^2}{} \,, &
\gamma_H =& -\frac12\,  \frac{\rd}{\rd \ln \mu} \coef{4}{H^2D^2}{} \,.
\end{align}

There is one redundant operator at dimension four, $\opr{4}{H^2\Box}{}$, which can be eliminated by integration by parts,
\begin{align}
\opr{4}{H^2\Box}{} &\equiv (H^\dagger D^2 H) + (D^2 H^\dagger H) = - 2 (D_\mu H^\dagger D^\mu H) = -2\ \op{4}{H^2D^2}{} \,.
\label{A.12}
\end{align}
The dimension-four operators are listed in Table~\ref{dim:4}.

\subsection{Dimension 6}

At dimension six, we use the operators in the Warsaw basis~\cite{Grzadkowski:2010es}, which are listed in Table~\ref{dim:6}. The redundant dimension-six operators are listed in Table~\ref{red:6}. They can be eliminated by field redefinitions and integration by parts. The operator relations for these are listed below, and extend previous results to include dimension-eight contributions to the reduction formula. The relations all respect the NDA power counting rules~\cite{Manohar:1983md,Jenkins:2013sda,Gavela:2016bzc}, which dictate the power of coupling constants in the various terms.

Note that the relations need not remain valid when multiplied by additional fields, because integration by parts has been used in their derivation. For example, one has the relation in Eq.~\eqref{A.12},
\begin{align}
H^\dagger D^2 H + D^2 H^\dagger H + 2 (D_\mu H^\dagger D^\mu H) &=0 \,,
\label{B.14}
\end{align}
but on multiplying by $(H^\dagger H)$,
\begin{align}
(H^\dagger H) \left[ H^\dagger D^2 H + D^2 H^\dagger H+ 2 (D_\mu H^\dagger D^\mu H) \right] &= \op{6}{H^4\Box}{} \,,
\label{B.15}
\end{align}
instead of zero.

\subsubsection{$H^4 D^2$}

The redundant $H^4 D^2$ operators  are related to the operators in the canonical basis as
\begin{align}
\opr{6}{H^4\Box}{} = & m_H^2 \op{4}{H^4}{}  + \left( -4 \lambda + 4 m_H^2\ \coef{6}{H^4\Box}{} -\frac12 m_H^2\ {\coef{6}{H^4D^2}{}} \right) \op{6}{H^6}{} \nn
& + \left( 6\ \ \coef{6}{H^6}{} -16 \lambda\ \coef{6}{H^4\Box}{} + 2\lambda\ {\coef{6}{H^4D^2}{}} \right) \op{8}{H^8}{} + \left( 8\ \ \coef{6}{H^4\Box}{}+{\coef{6}{H^4D^2}{}} \right) \op{8}{H^6 D^2}{(1)} \nn
& + 2\ \ {\coef{6}{H^4D^2}{}} \op{8}{H^6 D^2}{(2)} + 2\ \ \coef{6}{G^2H^2}{} \op{8}{G^2H^4}{(1)}
+ 2\ \ \coef{6}{W^2H^2}{} \op{8}{W^2H^4}{(1)} +  2\ \ \coef{6}{B^2H^2}{} \op{8}{B^2H^4}{(1)} \nn
& + 2\ \ {\coef{6}{WBH^2}{}} \op{8}{WBH^4}{(1)} \,,
\label{B.8}
\end{align}

\begin{table}
\begin{center}
\begin{minipage}[t]{8.75cm}
\vspace{-0.5cm}
\renewcommand{\arraystretch}{1.5}
\begin{align*}
\begin{array}{c|c}
\multicolumn{2}{c}{\bm{H^4 D^2} } \\
\hline
\opr{6}{H^4\Box}{} & (H^\dagger H) \left[ H^\dagger D^2 H + D^2  H^\dagger H\right]  \\
\opr{6}{H^4D^2}{(1)} & (H^\dagger H) (D_\mu H^\dagger D^\mu H) \\
\opr{6}{H^4D^2}{(2)}  & (H^\dagger D_\mu H) (H^\dagger D^\mu H) + (D_\mu H^\dagger H)(D^\mu H^\dagger H) \\
\opr{6}{H^4D^2}{(3)} & j_{\mu} j^{ \mu} \\
 \opr{6}{H^4D^2}{(4)} &j_{\mu}^a j^{a\, \mu} 
\end{array}
\end{align*}
\end{minipage}
%
\hspace{0.1cm}
\begin{minipage}[t]{6cm}
\vspace{-0.5cm}
\renewcommand{\arraystretch}{1.5}
\begin{align*}
\begin{array}{c|c}
\multicolumn{2}{c}{\bm{X H^2 D^2} } \\
\hline
\opr{6}{WH^2D^2}{(1)} & 
 j^{a \, \nu} D^\mu W^a_{\mu \nu} \\
\opr{6}{WH^2D^2}{(2)} & i  (D^\mu H)^\dagger \, \tau^a \, (D^\nu H) \,  W^a_{\mu \, \nu} \\
\opr{6}{BH^2D^2}{(1)} & 
 j^\nu \partial^\mu B_{\mu \nu}  \\
\opr{6}{BH^2D^2}{(2)} & i (D^\mu H)^\dagger \,  (D^\nu H) \,  B_{\mu \, \nu} \\
\end{array}
\end{align*}
\end{minipage}
\end{center}
\caption{\label{red:6} Redundant bosonic dimension-six operators in the SMEFT.}
\end{table}

\begin{align}
\opr{6}{H^4D^2}{(1)} = &  \frac12\ \op{6}{H^4\Box}{} - \frac12 \ \opr{6}{H^4\Box}{} \,,
\label{B.10}
\end{align}
\begin{align}
\opr{6}{H^4D^2}{(2)} = & -\op{6}{H^4\Box}{} -2\ \ \op{6}{H^4 D^2}{} \,,
\label{B.11}
\end{align}
\begin{align}
\opr{6}{H^4D^2}{(3)} = & \op{6}{H^4\Box}{} + 4\ \ \op{6}{H^4 D^2}{} \,,
\label{B.12}
\end{align}
\begin{align}
\opr{6}{H^4D^2}{(4)}=& 3\ \  \op{6}{H^4\Box}{} - 2\ \  \opr{6}{H^4\Box}{} \,.
\label{B.13}
\end{align}

\subsubsection{$X H^2 D^2$}

The redundant $X H^2 D^2$ operators are related to the operators in the canonical basis as
\begin{align}
\opr{6}{WH^2D^2}{(1)} =&  -g_2 m_H^2\ \op{4}{H^4}{} + \frac32 g_2\ \op{6}{H^4\Box}{}
\nn &
+ \left(4 g_2 \lambda - 4 g_2 m_H^2\ \coef{6}{H^4\Box}{}  +\frac34 g_2 m_H^2\ {\coef{6}{H^4D^2}{}} +
g_1  m_H^2\ {\coef{6}{WBH^2}{}} \right) \op{6}{H^6}{} \nn
&  + \left(-6 g_2\ \coef{6}{H^6}{} + 16 g_2 \lambda\ \coef{6}{H^4\Box}{} -3 g_2 \lambda\ {\coef{6}{H^4D^2}{}} - 4 g_1  \lambda\  {\coef{6}{WBH^2}{}} \right) \op{8}{H^8}{} \nn
& + \left(-8 g_2\ \coef{6}{H^4\Box}{} +\frac12 g_2\  {\coef{6}{H^4D^2}{}} + 6 g_1\  {\coef{6}{WBH^2}{}} \right) \op{8}{H^6D^2}{(1)} \nn
& + \left(- g_2\ {\coef{6}{H^4D^2}{}} + 4 g_1\  {\coef{6}{WBH^2}{}} \right) \op{8}{H^6D^2}{(2)}  - 2 g_2\ \coef{6}{G^2H^2}{} \op{8}{G^2H^4}{(1)}\nn
& + \left(-2 g_2\ \coef{6}{B^2H^2}{} - g_1\  {\coef{6}{WBH^2}{}} \right)  \op{8}{B^2H^4}{(1)} 
\nn &
+  \left( 2 g_1\  \coef{6}{W^2H^2}{}  -3 g_2\ {\coef{6}{WBH^2}{}}\right)  \op{8}{WBH^4}{(1)}\nn
&  - 8\ \ \coef{6}{W^2H^2}{}  \op{8}{WH^4D^2}{(1)} + 12\ \ {\coef{6}{WBH^2}{}}  \op{8}{BH^4D^2}{(1)}  \,,
\label{B.17}
\end{align}

\begin{align}
\opr{6}{WH^2D^2}{(2)} =&  -\frac12\ \opr{6}{WH^2D^2}{(1)} +\frac14 g_2\ \op{6}{W^2H^2}{} + \frac14 g_1\ \op{6}{WBH^2}{} \,,
\label{B.18}
\end{align}

\begin{align}
\opr{6}{BH^2D^2}{(1)} =&   \left(\frac14 g_1  m_H^2\ {\coef{6}{H^4D^2}{}} + \frac12 g_2 m_H^2\ {\coef{6}{WBH^2}{}} \right) \op{6}{H^6}{}  + \frac12 g_1\  \op{6}{H^4\Box}{} +2 g_1\ \op{6}{H^4 D^2}{} \nn
& + 
\left(- g_1   \lambda\ {\coef{6}{H^4D^2}{}} - 2 g_2\lambda\ {\coef{6}{WBH^2}{}} \right) \op{8}{H^8}{} 
\nn &
+ \left( \frac{3}{2} g_1 \   {\coef{6}{H^4D^2}{}} + 4 g_2\  {\coef{6}{WBH^2}{}} \right) \op{8}{H^6D^2}{(1)} 
 +  \left(g_1  \  {\coef{6}{H^4D^2}{}} +  g_2 \ {\coef{6}{WBH^2}{}} \right) \op{8}{H^6D^2}{(2)}
\nn &
-\frac12 g_2\ {\coef{6}{WBH^2}{}} \op{8}{W^2H^4}{(1)} +  \frac12 g_2\ {\coef{6}{WBH^2}{}} \op{8}{W^2H^4}{(3)}  
+ 2 g_1 \ \coef{6}{B^2H^2}{} \op{8}{B^2H^4}{(1)}
\nn
&  + 2 g_2 \ \coef{6}{B^2H^2}{} \op{8}{WBH^4}{(1)} +  4\ \ {\coef{6}{WBH^2}{}} \op{8}{WH^4D^2}{(1)} - 8 \ \ \coef{6}{B^2H^2}{} \op{8}{BH^4D^2}{(1)}  \,,
\label{B.19}
\end{align}

\begin{align}
\opr{6}{BH^2D^2}{(2)} =&  -\frac12\ \opr{6}{BH^2D^2}{(1)} +\frac14 g_1\  \op{6}{B^2H^2}{} + \frac14 g_2\ \op{6}{WBH^2}{} \,.
\label{B.20}
\end{align}

\subsection{Dimension 8}

The dimension-eight even-parity bosonic operator basis is that of Ref.~\cite{Murphy:2020rsh}. The operators are given in Table~\ref{dim:8}. The operator superscripts are not in consecutive order, e.g., $Q_{W^2H^2D^2}^{(n)}$, with $n=1,2,4$. This because the $n=3$ operator in Ref.~\cite{Murphy:2020rsh} has odd parity, and is not used in this paper. Redundant bosonic dimension-eight operators are given in Tables~\ref{red:81},~\ref{red:82}, and~\ref{red:83} and the relations used to eliminate these are listed below.

\subsubsection{$H^6 D^2$}

The redundant $H^6D^2$ operators are related to the operators in the canonical basis as
\begin{align}
\opr{8}{H^6\Box}{} &= m_H^2\ \op{6}{H^6}{} - 4 \lambda\ \op{8}{H^8}{} \,,
\label{B.22}
\end{align}
\begin{align}
\opr{8}{H^6D^2}{(1)} = & \frac12\ \op{8}{H^6D^2}{(1)} + \frac12\ \op{8}{H^6D^2}{(2)} \,,
\label{B.24}
\end{align}
\begin{align}
\opr{8}{H^6D^2}{(2)} = & -\frac12 m_H^2\ \op{6}{H^6}{} + 2 \lambda\ \op{8}{H^8}{} -2\ \  \op{8}{H^6D^2}{(1)} - \op{8}{H^6D^2}{(2)} \,,
\label{B.25}
\end{align}
\begin{align}
\opr{8}{H^6D^2}{(3)} = & -\frac12 m_H^2 \ \op{6}{H^6}{} + 2 \lambda \ \op{8}{H^8}{} -  \op{8}{H^6D^2}{(1)} \,,
\label{B.26}
\end{align}
\begin{align}
\opr{8}{H^6D^2}{(4)} = & \frac12 m_H^2 \ \op{6}{H^6}{} - 2 \lambda \ \op{8}{H^8}{} +3\  \ \op{8}{H^6D^2}{(1)} + 2\ \  \op{8}{H^6D^2}{(2)} \,,
\label{B.27}
\end{align}
\begin{align}
\opr{8}{H^6D^2}{(5)} = & \frac12 m_H^2 \ \op{6}{H^6}{} - 2 \lambda \ \op{8}{H^8}{} +3\  \ \op{8}{H^6D^2}{(1)} + 2\  \ \op{8}{H^6D^2}{(2)} \,,
\label{B.28}
\end{align}
\begin{align}
\opr{8}{H^6D^2}{(6)} = & \frac12 m_H^2 \ \op{6}{H^6}{} - 2 \lambda \ \op{8}{H^8}{} + 5\  \ \op{8}{H^6D^2}{(1)}  \,.
\label{B.29}
\end{align}

\subsubsection{$X H^4 D^2$}

The redundant $X H^4D^2$ operators are related to the operators in the canonical basis as
\begin{align}
\opr{8}{WH^4D^2}{(1)} = & 0 \,,
\label{B.31}
\end{align}
\begin{align}
\opr{8}{WH^4D^2}{(2)} = & \frac12 g_2 \ \op{8}{H^6D^2}{(1)} - \frac12 g_2 \ \op{8}{H^6D^2}{(2)} - \frac14 g_2 \ \op{8}{W^2H^4}{(1)} + \frac14 g_2 \ \op{8}{W^2H^4}{(3)} + 2 \ \ \op{8}{WH^4D^2}{(1)} \,,
\label{B.32}
\end{align}
\begin{align}
\opr{8}{WH^4D^2}{(3)} = & \frac14 g_2 m_H^2 \ \op{6}{H^6}{} - g_2 \lambda \ \op{8}{H^8}{} + \frac52 g_2 \ \op{8}{H^6D^2}{(1)}  - \frac12 g_2 \ \op{8}{W^2H^4}{(1)} - \frac12 g_1  \ \op{8}{WBH^4}{(1)} \nn
& + 2\  \ \op{8}{WH^4D^2}{(1)} \,,
\label{B.33}
\end{align}
\begin{align}
\opr{8}{WH^4D^2}{(4)} = 2\  \ \op{8}{WH^4D^2}{(3)}  \,,
\label{B.34}
\end{align}
\begin{align}
\opr{8}{WH^4D^2}{(5)} = & g_2 \ \op{8}{H^6D^2}{(1)} -  g_2 \ \op{8}{H^6D^2}{(2)} - \frac12 g_2 \ \op{8}{W^2H^4}{(1)} + \frac12 g_2 \ \op{8}{W^2H^4}{(3)}  \,,
\label{B.35}
\end{align}
\begin{align}
\opr{8}{WH^4D^2}{(6)} = -2\ \  \op{8}{WH^4D^2}{(3)}  \,,
\label{B.36}
\end{align}
\begin{align}
\opr{8}{WH^4D^2}{(7)} = & \frac12 g_2 m_H^2 \ \op{6}{H^6}{} - 2 g_2 \lambda \ \op{8}{H^8}{} + 5 g_2 \ \op{8}{H^6D^2}{(1)}  - g_2 \ \op{8}{W^2H^4}{(1)} -   g_1  \ \op{8}{WBH^4}{(1)} \nn
& + 8\  \ \op{8}{WH^4D^2}{(1)} \,,
\label{B.37}
\end{align}
\begin{align}
\opr{8}{WH^4D^2}{(8)} =  \op{8}{WH^4D^2}{(3)} \,,
\label{B.38}
\end{align}
\begin{align}
\opr{8}{WH^4D^2}{(9)} = & -\frac18 g_2 m_H^2 \ \op{6}{H^6}{} + \frac12 g_2 \lambda \ \op{8}{H^8}{} - g_2 \ \op{8}{H^6D^2}{(1)}  - \frac14 g_2  \ \op{8}{H^6D^2}{(2)}  \nn
& + \frac18 g_2 \ \op{8}{W^2H^4}{(1)} + \frac18 g_2 \ \op{8}{W^2H^4}{(3)} + \frac14  g_1  \ \op{8}{WBH^4}{(1)}  \,,
\label{B.39}
\end{align}
\begin{align}
\opr{8}{WH^4D^2}{(10)} = & \op{8}{WH^4D^2}{(3)}  \,,
\label{B.40}
\end{align}
\begin{align}
\opr{8}{WH^4D^2}{(11)} = & -\frac18 g_2 m_H^2 \ \op{6}{H^6}{} + \frac12 g_2 \lambda \ \op{8}{H^8}{} - \frac32 g_2 \ \op{8}{H^6D^2}{(1)}  + \frac14 g_2  \ \op{8}{H^6D^2}{(2)} 
+ \frac38 g_2 \ \op{8}{W^2H^4}{(1)}  \nn
& - \frac18 g_2 \ \op{8}{W^2H^4}{(3)}  + \frac14  g_1  \ \op{8}{WBH^4}{(1)} -2\  \ \op{8}{WH^4D^2}{(1)}  \,,
\label{B.41}
\end{align}
\begin{align}
\opr{8}{WH^4D^2}{(12)} = & -\frac12 g_2 \ \op{8}{W^2H^4}{(1)}  + \frac12 g_2 \ \op{8}{W^2H^4}{(3)} \,,
\label{B.42}
\end{align}
\begin{align}
\opr{8}{BH^4D^2}{(1)} = & -\frac18 g_1  m_H^2 \ \op{6}{H^6}{} + \frac12 g_1  \lambda \ \op{8}{H^8}{} - \frac34 g_1  \ \op{8}{H^6D^2}{(1)}  -  \frac12 g_1  \ \op{8}{H^6D^2}{(2)} \nn
& + \frac14 g_1  \ \op{8}{B^2H^4}{(1)}  + \frac14  g_2 \ \op{8}{WBH^4}{(1)} -  \op{8}{BH^4D^2}{(1)}  \,,
\label{B.43}
\end{align}
\begin{align}
\opr{8}{BH^4D^2}{(2)} = & \frac14 g_1  m_H^2 \ \op{6}{H^6}{} -  g_1  \lambda \ \op{8}{H^8}{} + \frac{3}{2}  g_1  \ \op{8}{H^6D^2}{(1)}  +  g_1   \ \op{8}{H^6D^2}{(2)} 
- \frac12 g_1  \ \op{8}{B^2H^4}{(1)} \nn
& - \frac12  g_2 \ \op{8}{WBH^4}{(1)} +2\   \ \op{8}{BH^4D^2}{(1)} = - 2\ \ \opr{8}{BH^4D^2}{(1)}  \,,
\label{B.44}
\end{align}
\begin{align}
\opr{8}{BH^4D^2}{(3)} = & 2 \ \ \opr{8}{BH^4D^2}{(1)} - \op{8}{BH^4D^2}{(1)} \,.
\end{align}

\subsubsection{$X^2 H^2 D^2$}

The redundant $X^2 H^2 D^2$ operators are related to the operators in the canonical basis as
\begin{align}
\opr{8}{G^2H^2D^2}{(1)} = \frac12 m_H^2 \ \op{6}{G^2H^2}{} -2 \lambda \ \op{8}{G^2H^4}{(1)} + 2 g_3 \ \op{8}{G^3H^2}{(1)} + \op{8}{G^2H^2D^2}{(2)} \,,
\label{B.44a}
\end{align}
\begin{align}
\opr{8}{G^2H^2D^2}{(2)} = \frac12 \ \opr{8}{G^2H^2D^2}{(1)} \,,
\label{B.45}
\end{align}
\begin{align}
\opr{8}{G^2H^2D^2}{(3)} = - \frac12 m_H^2 \ \op{6}{G^2H^2}{} + 2 \lambda \ \op{8}{G^2H^4}{(1)} - \op{8}{G^2H^2D^2}{(2)} \,,
\label{B.46}
\end{align}
\begin{align}
\opr{8}{G^2H^2D^2}{(4)} = \frac12 \ \opr{8}{G^2H^2D^2}{(3)} \,,
\label{B.47}
\end{align}
\begin{align}
\opr{8}{W^2H^2D^2}{(1)} =&  \frac12 m_H^2 \ \op{6}{W^2H^2}{} + \left(-2 \lambda + \frac12 g_2^2 \right) \op{8}{W^2H^4}{(1)} +\frac12  g_1 g_2  \ \op{8}{WBH^4}{(1)} 
+ 2 g_2 \ \op{8}{W^3H^2}{(1)} \nn
& + \op{8}{W^2H^2D^2}{(2)} - 2 g_2 \ \op{8}{WH^4D^2}{(1)}  \,,
\label{B.48}
\end{align}
\begin{align}
\opr{8}{W^2H^2D^2}{(2)} =&  \frac12  \ \opr{8}{W^2H^2D^2}{(1)}  \,,
\label{B.49}
\end{align}
\begin{align}
\opr{8}{W^2H^2D^2}{(3)} =& -\frac12 m_H^2 \ \op{6}{W^2H^2}{} +  2 \lambda \ \op{8}{W^2H^4}{(1)} - \op{8}{W^2H^2D^2}{(2)} \,,
\label{B.50}
\end{align}
\begin{align}
\opr{8}{W^2H^2D^2}{(4)} =&  \frac12  \ \opr{8}{W^2H^2D^2}{(3)}  \,,
\label{B.51}
\end{align}
\begin{align}
\opr{8}{W^2H^2D^2}{(5)} =& -\frac12 g_2^2 m_H^2 \ \op{6}{H^6}{} + 2 g_2^2 \lambda \ \op{8}{H^8}{} - 5 g_2^2 \ \op{8}{H^6D^2}{(1)} + g_2^2 \ \op{8}{W^2H^4}{(1)}
+  g_1 g_2   \ \op{8}{WBH^4}{(1)}  \nn
& + g_2  \ \op{8}{W^3H^2}{(1)} +  g_1    \ \op{8}{W^2BH^2}{(1)} -4 \  \ \op{8}{W^2H^2D^2}{(4)}
- 8 g_2   \ \op{8}{WH^4D^2}{(1)} \,,
\label{B.52}
\end{align}
\begin{align}
\opr{8}{W^2H^2D^2}{(6)} =&  \frac12  \ \opr{8}{W^2H^2D^2}{(5)}  \,,
\label{B.53}
\end{align}
\begin{align}
\opr{8}{B^2H^2D^2}{(1)} =&  \frac12  m_H^2 \ \op{6}{B^2H^2}{} + \left(-2 \lambda + \frac12 g_1^2  \right) \op{8}{B^2H^4}{(1)} + \frac12 g_1 g_2  \ \op{8}{WBH^4}{(1)}
+ \op{8}{B^2H^2D^2}{(2)}
\nn &
- 2 g_1  \ \op{8}{BH^4D^2}{(1)} \,,
\label{B.54}
\end{align}
\begin{align}
\opr{8}{B^2H^2D^2}{(2)} =&  \frac12  \ \opr{8}{B^2H^2D^2}{(1)}  \,,
\label{B.55}
\end{align}
\begin{align}
\opr{8}{B^2H^2D^2}{(3)} =& - \frac12  m_H^2 \ \op{6}{B^2H^2}{} +  2 \lambda \ \op{8}{B^2H^4}{(1)} - \op{8}{B^2H^2D^2}{(2)}   \,,
\label{B.56}
\end{align}
\begin{align}
\opr{8}{B^2H^2D^2}{(4)} =&  \frac12  \ \opr{8}{B^2H^2D^2}{(3)}  \,,
\label{B.57}
\end{align}
\begin{align}
\opr{8}{WBH^2D^2}{(1)} =&  \frac38 g_1 g_2  m_H^2 \ \op{6}{H^6}{} + \frac12 m_H^2 \ \op{6}{WBH^2}{} - \frac{3}{2} g_1 g_2  \lambda \ \op{8}{H^8}{} + \frac{5}{2} g_1 g_2  \ \op{8}{H^6 D^2}{(1)} \nn
& + \frac54 g_1 g_2  \ \op{8}{H^6 D^2}{(2)}  - \frac18  g_1 g_2  \ \op{8}{W^2 H^4}{(1)} 
+  \frac18  g_1 g_2  \ \op{8}{W^2 H^4}{(3)} -\frac14 g_1 g_2  \ \op{8}{B^2 H^4}{(1)} \nn
& - \left(2 \lambda + \frac14 g_2^2 \right) \op{8}{WBH^4}{(1)} +g_2 \ \op{8}{W^2 B H^2}{(1)}  + \op{8}{W B H^2 D^2}{(1)} +   g_1   \ \op{8}{W H^4 D^2}{(1)}
\nn &
+ 3 g_2  \ \op{8}{B H^4 D^2}{(1)} \,,
\label{B.58}
\end{align}
\begin{align}
\opr{8}{WBH^2D^2}{(2)} =&  \frac12  \ \opr{8}{WBH^2D^2}{(1)}  \,,
\label{B.59}
\end{align}
\begin{align}
\opr{8}{WBH^2D^2}{(3)} =&  -\frac18 g_1 g_2  m_H^2 \ \op{6}{H^6}{} - \frac12 m_H^2 \ \op{6}{WBH^2}{} +  \frac12 g_1 g_2  \lambda \ \op{8}{H^8}{} -
 \frac12 g_1 g_2  \ \op{8}{H^6 D^2}{(1)} \nn
 & - \frac34 g_1 g_2  \ \op{8}{H^6 D^2}{(2)}  - \frac18  g_1 g_2  \ \op{8}{W^2 H^4}{(1)} 
+  \frac18  g_1 g_2  \ \op{8}{W^2 H^4}{(3)} + \frac14 g_1 g_2  \ \op{8}{B^2 H^4}{(1)}  \nn
& + \left(2 \lambda + \frac14 g_2^2 \right) \op{8}{WBH^4}{(1)} - g_2 \ \op{8}{W^2 B H^2}{(1)}  - \op{8}{W B H^2 D^2}{(1)} +  g_1   \ \op{8}{W H^4 D^2}{(1)}
\nn &
- 3 g_2  \ \op{8}{B H^4 D^2}{(1)} \,,
\label{B.60}
\end{align}
\begin{align}
\opr{8}{WBH^2D^2}{(4)} =&  \frac12  \ \opr{8}{WBH^2D^2}{(3)}  \,,
\label{B.61}
\end{align}
\begin{align}
\opr{8}{WBH^2D^2}{(5)} =&  \frac18 g_1 g_2  m_H^2 \ \op{6}{H^6}{} - \frac12 m_H^2 \ \op{6}{WBH^2}{} -  \frac12 g_1 g_2  \lambda \ \op{8}{H^8}{} +
\frac12 g_1 g_2  \ \op{8}{H^6 D^2}{(1)}
\nn &
+ \frac34 g_1 g_2  \ \op{8}{H^6 D^2}{(2)} 
 + \frac18  g_1 g_2  \ \op{8}{W^2 H^4}{(1)} 
-  \frac18  g_1 g_2  \ \op{8}{W^2 H^4}{(3)} - \frac14 g_1 g_2  \ \op{8}{B^2 H^4}{(1)} 
\nn &
+ \left(2 \lambda - \frac14 g_2^2 \right)  \op{8}{WBH^4}{(1)} 
 + g_2 \ \op{8}{W^2 B H^2}{(1)}  -  \op{8}{W B H^2 D^2}{(1)} -   g_1   \ \op{8}{W H^4 D^2}{(1)}
\nn &
 + 3 g_2  \ \op{8}{B H^4 D^2}{(1)} \,,
\label{B.62}
\end{align}
\begin{align}
\opr{8}{WBH^2D^2}{(6)} =&  \frac12  \ \opr{8}{WBH^2D^2}{(5)}  \,.
\label{B.63}
\end{align}

\subsubsection{$H^4 D^4$}

The redundant $H^4 D^4$ operators are related to the operators in the canonical basis as
\begin{align}
\opr{8}{H^4 \Box^2}{(1)} = & \opr{8}{H^4 \Box^2}{(2)} =  \frac12 \ \opr{8}{H^4 \Box^2}{(3)} = \frac14 m_H^4 \ \op{4}{H^4}{} -2 \lambda m_H^2 \ \op{6}{H^6}{} +4 \lambda^2 \ \op{8}{H^8}{} \,,
\end{align}
\begin{align}
\opr{8}{H^4D^2\Box}{(1)} = &  -\frac12 m_H^4 \ \op{4}{H^4}{} + 2 \lambda m_H^2  \ \op{6}{H^6}{} + \frac12 m_H^2 \ \op{6}{H^4\Box}{}  -4 \lambda  \ \op{8}{H^6D^2}{(1)}  \,,
\end{align}
\begin{align}
\opr{8}{H^4D^2\Box}{(2)} = &  m_H^2 \ \op{6}{H^4 D^2}{} - 2 \lambda \ \op{8}{H^6D^2}{(1)} - 2 \lambda \ \op{8}{H^6D^2}{(2)} \,,
\end{align}
\begin{align}
\opr{8}{H^4D^4}{(1)} = &  m_H^2 \left(-\frac12 \lambda + \frac1{16} g_1^2  + \frac{1}{16} g_2^2 \right) \op{6}{H^6}{} 
 +  \lambda\left( 2\lambda  - \frac14 g_1^2  - \frac14 g_2^2 \right) \op{8}{H^8}{} 
+ \op{8}{H^4D^4}{(3)} 
 \nn
 & + \left(-\lambda + \frac38 g_1^2 + \frac58 g_2^2 \right) \op{8}{H^6D^2}{(1)} + \frac14 g_1^2  \ \op{8}{H^6D^2}{(2)}
 + g_2 \ \op{8}{WH^4D^2}{(1)} +  g_1    \ \op{8}{BH^4D^2}{(1)} \,,
\end{align}
\begin{align}
\opr{8}{H^4D^4}{(2)} = &  m_H^2 \left(-\frac12 \lambda - \frac1{16} g_1^2  \right) \op{6}{H^6}{} 
 +  \lambda\left( 2\lambda  + \frac14 g_1^2 \right) \op{8}{H^8}{}  + \left(-\lambda - \frac38 g_1^2 - \frac18 g_2^2 \right) \op{8}{H^6D^2}{(1)}  \nn
 & + \left(-\frac14 g_1^2 +\frac18 g_2^2 \right) \op{8}{H^6D^2}{(2)}
 + \op{8}{H^4D^4}{(1)} +\frac1{16}  g_2^2 \ \op{8}{W^2H^4}{(1)}  +\frac1{16}  g_2^2 \ \op{8}{W^2H^4}{(3)}  \nn
 &
 +\frac14 g_1^2  \ \op{8}{B^2H^4}{(1)}
 + \frac38 g_1 g_2   \ \op{8}{WBH^4}{(1)}
 -  g_1    \ \op{8}{BH^4D^2}{(1)} \,,
\end{align}
\begin{align}
\opr{8}{H^4D^4}{(3)} = &  m_H^2 \left(- \lambda - \frac18 g_1^2  -\frac14 g_2^2 \right) \op{6}{H^6}{} 
 +  \lambda\left( 4 \lambda  + \frac12  g_1^2  +g_2^2 \right) \op{8}{H^8}{}  \nn
 &  + \left(-2 \lambda - \frac34 g_1^2 - \frac94 g_2^2 \right) \op{8}{H^6D^2}{(1)}  + \left(-\frac12 g_1^2  - \frac14 g_2^2 \right) \op{8}{H^6D^2}{(2)}
 + 2\ \ \op{8}{H^4D^4}{(2)}  \nn
 &+\frac3{8}  g_2^2  \ \op{8}{W^2H^4}{(1)}  - \frac1{8} g_2^2  \ \op{8}{W^2H^4}{(3)} 
  + \frac14 g_1 g_2   \ \op{8}{WBH^4}{(1)}
 - 2 g_2  \ \op{8}{WH^4D^2}{(1)} \,,
\end{align}
\begin{align}
\opr{8}{H^4D^4}{(4)} = &  -\frac14 m_H^4 \ \op{4}{H^4}{} + m_H^2 \left(\frac12 \lambda + \frac1{16} g_1^2  \right) \op{6}{H^6}{} +\frac12 m_H^2 \ \op{6}{H^4\Box}{}
 +  \lambda\left( 2 \lambda  -\frac14  g_1^2 \right) \op{8}{H^8}{}  \nn
 &  + \left(-3 \lambda + \frac3{8} g_1^2  + \frac18 g_2^2 \right) \op{8}{H^6D^2}{(1)}  + \left(\frac14 g_1^2 - \frac18 g_2^2 \right) \op{8}{H^6D^2}{(2)}
  \nn
 &-\frac1{16}g_2^2   \ \op{8}{W^2H^4}{(1)}  + \frac1{16}g_2^2   \ \op{8}{W^2H^4}{(3)} - \frac18 g_1^2  \ \op{8}{B^2H^4}{(1)}  
  - \frac18 g_1 g_2  \ \op{8}{WBH^4}{(1)}
 +  g_1 \ \op{8}{BH^4D^2}{(1)}
\nn &
- \op{8}{H^4D^4}{(1)}  - \op{8}{H^4D^4}{(2)} + \op{8}{H^4D^4}{(3)} \,,
\end{align}
\begin{align}
\opr{8}{H^4D^4}{(5)} = & \opr{8}{H^4D^4}{(4)} + \frac18 m_H^2 g_2^2 \ \op{6}{H^6}{} - \frac12 \lambda g_2^2 \ \op{8}{H^8}{} +  g_2^2  \ \op{8}{H^6D^2}{(1)}  +\frac14 g_2^2 \ \op{8}{H^6D^2}{(2)}
 -\frac1{8}g_2^2   \ \op{8}{W^2H^4}{(1)} \nn
 &  - \frac1{8}g_2^2   \ \op{8}{W^2H^4}{(3)}   - \frac14 g_1 g_2  \ \op{8}{WBH^4}{(1)} + g_2  \ \op{8}{WH^4D^2}{(1)}
 -  g_1  \ \op{8}{BH^4D^2}{(1)} \,,
\end{align}
\begin{align}
\opr{8}{H^4D^4}{(6)} = &  \frac14 m_H^4 \ \op{4}{H^4}{} + m_H^2 \left(-\frac32 \lambda - \frac1{16} g_1^2  \right) \op{6}{H^6}{} + m_H^2 \ \op{6}{H^4 D^2}{}
 +  \lambda\left( 2 \lambda  + \frac14 g_1^2 \right) \op{8}{H^8}{}  \nn
 &  + \left(- \lambda - \frac38 g_1^2 - \frac18 g_2^2 \right) \op{8}{H^6D^2}{(1)}  + \left(-2\lambda - \frac14 g_1^2 + \frac18 g_2^2 \right) \op{8}{H^6D^2}{(2)} \nn
&  + \op{8}{H^4D^4}{(1)}  - \op{8}{H^4D^4}{(2)} - \op{8}{H^4D^4}{(3)}  +\frac1{16}g_2^2   \ \op{8}{W^2H^4}{(1)}  - \frac1{16}g_2^2   \ \op{8}{W^2H^4}{(3)} + \frac18 g_1^2  \ \op{8}{B^2H^4}{(1)}  \nn
&  + \frac18 g_1 g_2 \ \op{8}{WBH^4}{(1)}   -  g_1   \ \op{8}{BH^4D^2}{(1)} \,,
\end{align}
\begin{align}
\opr{8}{H^4D^4}{(7)} = & \opr{8}{H^4D^4}{(6)}  + m_H^2 \left( \frac18 g_1^2 + \frac1{16} g_2^2 \right) \op{6}{H^6}{}  +  \lambda\left( -\frac12 g_1^2  -\frac14 g_2^2 \right) \op{8}{H^8}{}  \nn
 &  + \left(\frac34  g_1^2 + \frac12 g_2^2 \right) \op{8}{H^6D^2}{(1)}  + \left(\frac12 g_1^2 + \frac18 g_2^2 \right) \op{8}{H^6D^2}{(2)} \nn
& - \frac1{16}g_2^2   \ \op{8}{W^2H^4}{(1)}  - \frac1{16}g_2^2   \ \op{8}{W^2H^4}{(3)} - \frac14 g_1^2   \ \op{8}{B^2H^4}{(1)}  
  - \frac38 g_1 g_2  \ \op{8}{WBH^4}{(1)} \nn
&  +  g_1  \ \op{8}{BH^4D^2}{(1)} \,,
\end{align}
\begin{align}
\opr{8}{H^4D^4}{(8)} = &  \frac14 m_H^4 \ \op{4}{H^4}{} + m_H^2 \left(-\frac12 \lambda - \frac1{16} g_1^2-\frac18 g_2^2 \right) \op{6}{H^6}{} -\frac12 m_H^2 \ \op{6}{H^4\Box}{}- m_H^2 \ \op{6}{H^4 D^2}{} \nn
& +  \lambda\left( -2 \lambda  + \frac14 g_1^2  +\frac12 g_2^2 \right) \op{8}{H^8}{}   + \left(5 \lambda - \frac38 g_1^2  - \frac98 g_2^2 \right) \op{8}{H^6D^2}{(1)}  \nn
& + \left(2\lambda - \frac14 g_1^2 - \frac18 g_2^2 \right) \op{8}{H^6D^2}{(2)}   - \op{8}{H^4D^4}{(1)}  + \op{8}{H^4D^4}{(2)} - \op{8}{H^4D^4}{(3)}  +\frac3{16}g_2^2   \ \op{8}{W^2H^4}{(1)}  \nn
& + \frac1{16}g_2^2   \ \op{8}{W^2H^4}{(3)} + \frac18 g_1^2  \ \op{8}{B^2H^4}{(1)}  
  + \frac38 g_1 g_2  \ \op{8}{WBH^4}{(1)}   - g_2  \ \op{8}{WH^4D^2}{(1)} \,,
\end{align}
\begin{align}
\opr{8}{H^4D^4}{(9)} = & \opr{8}{H^4D^4}{(8)}  + \frac1{16} m_H^2 g_2^2 \ \op{6}{H^6}{}-\frac14   \lambda g_2^2  \ \op{8}{H^8}{}   + \frac34 g_2^2  \ \op{8}{H^6D^2}{(1)} -\frac18 g_2^2 \ \op{8}{H^6D^2}{(2)}  \nn
&
 - \frac3{16}g_2^2   \ \op{8}{W^2H^4}{(1)} 
+ \frac1{16}g_2^2   \ \op{8}{W^2H^4}{(3)} - \frac18 g_1 g_2 \ \op{8}{WBH^4}{(1)}  + g_2  \ \op{8}{WH^4D^2}{(1)} \,.
\end{align}

\begin{table}
\begin{center}
\begin{minipage}[t]{10.25cm}
\vspace{-0.5cm}
\renewcommand{\arraystretch}{1.5}
\begin{align*}
\begin{array}{c|c}
\multicolumn{2}{c}{\bm{H^6 D^2} } \\
\hline
\opr{8}{H^6\Box}{} & (H^\dagger H)^2 (H^\dagger D^2 H +D^2 H^\dagger H) \\
\opr{8}{H^6D^2}{(1)} & (H^\dagger H)\left[(D_\mu H^\dagger H) (H^\dagger D_\mu H) \right] \\
\opr{8}{H^6D^2}{(2)} & (H^\dagger H)\left[(D_\mu H^\dagger H) (D^\mu H^\dagger H) + (H^\dagger D_\mu H) (H^\dagger D^\mu H) \right] \\
\opr{8}{H^6D^2}{(3)} &  
(H^\dagger H) r_\mu r^\mu  \\
\opr{8}{H^6D^2}{(4)} & (H^\dagger H)  j_{\mu}  j_{\mu}  \\
\opr{8}{H^6D^2}{(5)}& (H^\dagger \tau^a H)  j_{\mu}  j^a_{\mu}  \\
\opr{8}{H^6D^2}{(6)} & (H^\dagger H)  j^a_{\mu}  j^a_{\mu} 
\end{array}
\end{align*}
\end{minipage}
%
\begin{minipage}[t]{10.85cm}
\renewcommand{\arraystretch}{1.5}
\begin{align*}
\begin{array}{c|c}
\multicolumn{2}{c}{\bm{H^4 D^4} } \\
\hline
\opr{8}{H^4 \Box^2}{(1)}  & (H^\dagger H)(D^2 H^\dagger D^2 H) \\
\opr{8}{H^4 \Box^2}{(2)} & (H^\dagger D^2 H)(D^2 H^\dagger H) \\
\opr{8}{H^4 \Box^2}{(3)}  & (H^\dagger D^2 H) (H^\dagger D^2 H)  + (D^2 H^\dagger H) (D^2 H^\dagger H) \\
\opr{8}{H^4D^2\Box}{(1)}  & (D_\mu H^\dagger D_\mu H) (H^\dagger D^2 H + D^2 H^\dagger H) \\
\opr{8}{H^4D^2\Box}{(2)} & (H^\dagger D_\mu H)(D_\mu H^\dagger D^2 H) + (D_\mu H^\dagger H) (D^2 H^\dagger D_\mu H) \\
\opr{8}{H^4D^4}{(1)} &  (H^\dagger H)(D_\mu D_\nu H^\dagger   D_\mu D_\nu H) \\
\opr{8}{H^4D^4}{(2)} & (D_\mu D_\nu H^\dagger H) (H^\dagger D_\mu D_\nu H) \\
\opr{8}{H^4D^4}{(3)} & (D_\mu D_\nu H^\dagger H)(D_\mu D_\nu H^\dagger H) + (H^\dagger D_\mu D_\nu H)(H^\dagger D_\mu D_\nu H) \\
\opr{8}{H^4D^4}{(4)}  & (H^\dagger  D_\mu D_\nu H )(D_\mu H^\dagger D_\nu H) + (D_\mu D_\nu H^\dagger  H)(D_\nu H^\dagger D_\mu H) \\
\opr{8}{H^4D^4}{(5)}  & (H^\dagger  D_\nu D_\mu H )(D_\mu H^\dagger D_\nu H) + (D_\nu D_\mu H^\dagger  H)(D_\nu H^\dagger D_\mu H) \\
\opr{8}{H^4D^4}{(6)} & (D_\mu H^\dagger  D_\mu D_\nu H )(H^\dagger D_\nu H)+ (D_\mu D_\nu H^\dagger  D_\mu H) (D_\nu H^\dagger H) \\
\opr{8}{H^4D^4}{(7)} & (D_\nu H^\dagger  D_\mu D_\nu H )(H^\dagger D_\mu H)+ (D_\mu D_\nu H^\dagger  D_\nu H) (D_\mu H^\dagger H) \\
\opr{8}{H^4D^4}{(8)} & (D_\mu H^\dagger  D_\mu D_\nu H )(D_\nu H^\dagger H)+ (D_\mu D_\nu H^\dagger  D_\mu H) ( H^\dagger  D_\nu H) \\
\opr{8}{H^4D^4}{(9)} & (D_\nu H^\dagger  D_\mu D_\nu H )(D_\mu H^\dagger H)+ (D_\mu D_\nu H^\dagger  D_\nu H) ( H^\dagger  D_\mu H) \\
\end{array}
\end{align*}
\end{minipage}
\end{center}
\caption{\label{red:81} Redundant even-parity bosonic dimension-eight operators in the SMEFT, part~1.}
\end{table}

\begin{table}
\begin{center}
\begin{minipage}[t]{11.5cm}
\vspace{-0.5cm}
\renewcommand{\arraystretch}{1.5}
\begin{align*}
\begin{array}{c|c}
\multicolumn{2}{c}{\bm{X H^4 D^2} } \\
\hline
\opr{8}{WH^4D^2}{(1)} & 
r^{a\,\mu} r^\nu W^a_{\mu \nu}   \\
\opr{8}{WH^4D^2}{(2)} & 
r^{a\,\mu} j^\nu W^a_{\mu \nu}  \\
\opr{8}{WH^4D^2}{(3)}  &  
 j^{a\,\mu} r^\nu W^a_{\mu \nu}\\
\opr{8}{WH^4D^2}{(4)}  & 
j^{a\,\mu} j^\nu W^a_{\mu \nu} \\
\opr{8}{WH^4D^2}{(5)}  & 
\epsilon^{abc} r^{a\,\mu} r^{b\, \nu} W^{c}_{\mu\nu}  \\
\opr{8}{WH^4D^2}{(6)} & 
\epsilon^{abc} r^{a\,\mu} j^{b\, \nu} W^{c}_{\mu\nu}  \\
\opr{8}{WH^4D^2}{(7)}  &
\epsilon^{abc} j^{a\,\mu} j^{b\, \nu} W^{c}_{\mu\nu}  \\
\opr{8}{WH^4D^2}{(8)} & (D^\mu H^\dagger \tau^a H) (H^\dagger D^\nu H) W^a_{\mu \nu} + (H^\dagger \tau^a D^\mu  H) (D^\nu H^\dagger  H) W^a_{\mu \nu}  \\
\opr{8}{WH^4D^2}{(9)}  & i (D^\mu H^\dagger \tau^a H) (H^\dagger D^\nu H) W^a_{\mu \nu} - i (H^\dagger \tau^a D^\mu  H) (D^\nu H^\dagger  H) W^a_{\mu \nu}  \\
\opr{8}{WH^4D^2}{(10)} & (H^\dagger D^\mu H) (H^\dagger \tau^a D^\nu H) W^a_{\mu \nu} + (D^\mu  H^\dagger H) (D^\nu  H^\dagger \tau^a H) W^a_{\mu \nu}  \\
\opr{8}{WH^4D^2}{(11)} & i  (H^\dagger D^\mu H) (H^\dagger \tau^a D^\nu H) W^a_{\mu \nu} - i  (D^\mu  H^\dagger H) (D^\nu  H^\dagger \tau^a H) W^a_{\mu \nu}   \\
\opr{8}{WH^4D^2}{(12)} &  \epsilon_{abc} (H^\dagger \tau^a H) (H^\dagger \tau^b D_\mu D_\nu H + D_\mu D_\nu H^\dagger \tau^b  H) W^c_{\mu \nu} \\
\opr{8}{BH^4D^2}{(1)} & i(D^\mu H^\dagger H)(H^\dagger D^\nu H) B_{\mu \nu}   \\
\opr{8}{BH^4D^2}{(2)} & 
j^\mu r^\nu B_{\mu \nu}  \\
\opr{8}{BH^4D^2}{(3)} & i (H^\dagger \tau^a H)    \, (D^\mu H^\dagger \tau_a D^\nu H) B_{\mu \nu}  \\
\end{array}
\end{align*}
\end{minipage}
\end{center}
\caption{\label{red:82} Redundant bosonic dimension-eight operators in the SMEFT, part~2.}
\end{table}

\begin{table}
\begin{center}
\begin{minipage}[t]{6cm}
\vspace{-0.5cm}
\renewcommand{\arraystretch}{1.5}
\begin{align*}
\begin{array}{c|c}
\multicolumn{2}{c}{\bm{X^2 H^2 D^2} } \\
\hline
\opr{8}{G^2H^2D^2}{(1)} & (H^\dagger H) D_\mu G^{\mathscr{A}}_{\nu \sigma} D^\mu G^{\mathscr{A} \nu \sigma}   \\
\opr{8}{G^2H^2D^2}{(2)} & (H^\dagger H) D_\mu G^{\mathscr{A}}_{\nu \sigma} D^\nu G^{\mathscr{A} \mu \sigma} \\
\opr{8}{G^2H^2D^2}{(3)}  & 
r^\mu D_\mu G^{\mathscr{A}}_{\nu \sigma} G^{\mathscr{A}\nu \sigma}   \\
\opr{8}{G^2H^2D^2}{(4)}  & 
r^\mu D_\nu G^{\mathscr{A}}_{\mu \sigma} G^{\mathscr{A}\nu \sigma}   \\
\opr{8}{W^2H^2D^2}{(1)} & (H^\dagger H) D_\mu W^a_{\nu \sigma} D^\mu W^{a\, \nu \sigma}   \\
\opr{8}{W^2H^2D^2}{(2)} & (H^\dagger H) D_\mu W^a_{\nu \sigma} D^\nu W^{a\, \mu \sigma}  \\
\opr{8}{W^2H^2D^2}{(3)}  & 
r^\mu D_\mu W^a_{\nu \sigma} W^{a\, \nu \sigma}     \\
\opr{8}{W^2H^2D^2}{(4)} & 
r^\mu D_\nu W^a_{\mu \sigma} W^{a\, \nu \sigma}  \\
\opr{8}{W^2H^2D^2}{(5)}  & 
i \epsilon^{abc} j^{a \mu} W^{b\, \nu \sigma}  D_\mu W^c_{\nu \sigma}   \\
\opr{8}{W^2H^2D^2}{(6)}  & 
 i \epsilon^{abc} j^{a\mu}  W^{b\, \nu \sigma}  D_\nu W^c_{\mu \sigma}\\
\opr{8}{B^2H^2D^2}{(1)} & (H^\dagger H) D_\mu B_{\nu \sigma} D^\mu B^{\nu \sigma}  \\
\opr{8}{B^2H^2D^2}{(2)} & (H^\dagger H) D_\mu B_{\nu \sigma} D^\nu B^{\mu \sigma}  \\
\opr{8}{B^2H^2D^2}{(3)} &
r^\mu \partial_\mu B_{\nu \sigma} B^{\nu \sigma}  \\
\opr{8}{B^2H^2D^2}{(4)}  & 
r^\mu \partial_\nu B_{\mu \sigma} B^{\nu \sigma}  \\
\opr{8}{WBH^2D^2}{(1)} &  (H^\dagger \tau^a H) \partial_\mu B_{\nu \sigma} D^\mu W^{a\, \nu \sigma}   \\
\opr{8}{WBH^2D^2}{(2)} & (H^\dagger \tau^a H) \partial_\mu B_{\nu \sigma} D^\nu W^{a\, \mu \sigma}  \\
\opr{8}{WBH^2D^2}{(3)} & 
r^{a \mu} \partial_\mu B_{\nu \sigma} W^{a\, \nu \sigma}   \\
\opr{8}{WBH^2D^2}{(4)} & 
r^{a \mu} \partial_\nu B_{\mu \sigma} W^{a\, \nu \sigma}   \\
\opr{8}{WBH^2D^2}{(5)} & 
r^{a\mu} D_\mu W^a_{\nu \sigma} B^{\nu \sigma}   \\
\opr{8}{WBH^2D^2}{(6)}  & 
r^{a\mu} D_\nu W^a_{\mu \sigma} B^{\nu \sigma}   \\
\end{array}
\end{align*}
\end{minipage}
\end{center}
\caption{\label{red:83} Redundant bosonic dimension-eight operators in the SMEFT, part~3.}
\end{table}

\section{Renormalization Group Evolution in the SMEFT to Dimension Eight}\label{sec:RGEresults}

This appendix lists the renormalization group equations up to dimension eight. The terms included in the RGEs are those that depend on the coefficients that are included in our initial Lagrangian in Eq.~\eqref{eq:Lagr} through the metric and the potential. The dimension-two coefficient $m_H^2$ and dimension-four coefficients $g_1, g_2, g_3, \lambda$ are included, as are the dimension-six coefficients
\begin{align}
& \coef{6}{H^6}{},\ \coef{6}{H^4\Box}{},\ \coef{6}{H^4D^2}{},\ \coef{6}{G^2H^2}{},\ \coef{6}{W^2H^2}{},\ \coef{6}{B^2H^2}{},\ {\coef{6}{WBH^2}{}},
\end{align}
and the dimension-eight coefficients
\begin{align}
\coef{8}{H^8}{},\  \coef{8}{H^6D^2}{(1)},\ \coef{8}{H^6D^2}{(2)},\ \coef{8}{G^2H^4}{(1)}, \ \coef{8}{W^2H^4}{(1)},\ \coef{8}{W^2H^4}{(3)},\ \coef{8}{B^2H^4}{(1)},\ \coef{8}{WBH^4}{(1)}\,.
\label{BB.2}
\end{align}
Unless otherwise specified, only the dimension-eight terms in the RGE are listed here. The full RGE are given by adding the dimension-four and dimension-six terms given previously in Refs.~\cite{Jenkins:2013zja,Jenkins:2013wua,Alonso:2013hga}.

In the RGEs, we follow the notation of Refs.~\cite{Jenkins:2013zja,Jenkins:2013wua,Alonso:2013hga}, and use a dot over the coefficient to denote $16 \pi^2 \mu \tfrac{\rd}{\rd \mu}$. Consequently, the anomalous dimensions are all multiplied by $16\pi^2$.

\subsection{Field Anomalous Dimensions}

The field anomalous dimensions to dimension-eight in the gauge used in this paper are:
\begin{subequations}\label{C.3}
\begin{align}
\gamma_H =& -g_1^2 - 3 g_2^2 + 3 m_H^2\ \coef{6}{H^4\Box}{} \,, \\
\gamma_G =& -11 g_3^2 - 4 m_H^2\ \coef{6}{G^2H^2}{} \,, \\
\gamma_W =& -\frac{43}{6} g_2^2 - 4 m_H^2\ \coef{6}{W^2H^2}{} \,, \\
\gamma_B =& \frac{1}{6} g_1^2 - 4 m_H^2\ \coef{6}{B^2H^2}{} \,.
\end{align}
\end{subequations}
The gauge $\beta$-functions do not include fermionic contributions. In Feynman gauge, the $H$ anomalous dimensions of Refs.~\cite{Jenkins:2013zja,Jenkins:2013wua,Alonso:2013hga} is
\begin{align}
\gamma_H =& -\frac12 g_1^2 - \frac32 g_2^2 +  m_H^2 \left(2 \ \ \coef{6}{H^4\Box}{} - \coef{6}{H^4D^2}{}  \right) \,. 
\end{align}
The gauge field anomalous dimensions are the same as Eq.~\eqref{C.3} if the fermions contributions are dropped.

\subsection{Dimension 0}

The RGE for the cosmological constant is
\begin{align}
\dot \Lambda &=\frac12 m_H^4 \,,
\end{align}
and is entirely dimension four. The dimension-six and dimension-eight terms vanish.

\subsection{Dimension 2}

The RGE for the Higgs mass is
\begin{align}
\dot m_H^2 &= 
m_H^2  \biggl\{ 12 \lambda - \frac32 g_1^2 - \frac92 g_2^2  \biggr\} +
m_H^4  \biggl\{ -4\ \ \coef{6}{H^2\Box}{} + 2\ \ \coef{6}{H^2D^2}{}  \biggr\} \,,
\end{align}
and has only dimension-four and dimension-six contributions given in Refs.~\cite{Jenkins:2013wua,Jenkins:2013zja,Alonso:2013hga}. The dimension-eight contributions vanish.

\subsection{Dimension 4}

The RGE for the Higgs self-coupling is
\begin{align}
\dot \lambda = &  
\biggl\{ 24 \lambda^2 - \lambda \left(3 g_1^2 + 9 g_2^2 \right) + \frac38 g_1^4 + \frac34 g_1^2 g_2^2 + \frac98 g_2^4 \biggr\} \nn
& + m_H^2 \biggl\{ \coef{6}{H^6}{} + \left(-32 \lambda + \frac{10}{3} g_2^2 \right)\coef{6}{H^4\Box}{} + \left(12 \lambda +\frac32 g_1^2-\frac32 g_2^2 \right) \coef{6}{H^4D^2}{}  + 9 g_2^2\ \coef{6}{W^2H^2}{} \nn
& \qquad \qquad + 3 g_1^2 \ \coef{6}{B^2H^2}{} + 3 g_1 g_2 \ \coef{6}{WBH^2}{} \biggr\} \nn
&+ m_H^4 \biggl\{ 6 \left(\coef{6}{H^4\Box}{}\right)^2 + \frac{5}{4} \left(\coef{6}{H^4D^2}{}\right)^2 - 6\ \ \coef{6}{H^4\Box}{} \coef{6}{H^4D^2}{} 
-32 \left(\coef{6}{G^2H^2}{}\right)^2  \nn
& \qquad \qquad -12 \left(\coef{6}{W^2H^2}{}\right)^2 -4 \left(\coef{6}{B^2H^2}{}\right)^2  + 6 \left( \coef{6}{WBH^2}{}\right)^2 -4\ \ \coef{8}{H^6D^2}{(1)} + 2\ \ \coef{8}{H^6D^2}{(2)}
 \biggr\} \,.
\end{align}
The dimension-four and dimension-six contributions were given in Refs.~\cite{Jenkins:2013wua,Jenkins:2013zja,Alonso:2013hga}. The dimension-eight terms are proportional to $m_H^4$.

%
%
\subsection{Dimension 6}

The dimension-eight RGEs for the dimension-six coefficients in the SMEFT Lagrangian are listed below. The contributions are all of order $m_H^2/M^4$ in the SMEFT power counting. The dimension-six contributions are given in Refs.~\cite{Jenkins:2013wua,Jenkins:2013zja,Alonso:2013hga}, \emph{and not included below.}

\subsubsection{$H^6$}
The RGE for the $H^6$ coupling is
\begin{align}
\dcoef{6}{H^6}{} = m_H^2 &\biggl\{ -120\ \ \coef{6}{H^6}{} \coef{6}{H^4\Box}{} + 27\ \ \coef{6}{H^6}{} \coef{6}{H^4D^2}{} + \left( 320 \lambda + \frac{10}{3} g_1^2 -10 g_2^2 \right) \left(\coef{6}{H^4\Box}{}\right)^2  \nn
& + \left( \frac{35}{2} \lambda + \frac{37}{48} g_1^2 - \frac{5}{16} g_2^2 \right) \left( \coef{6}{H^4D^2}{} \right)^2 
+ \left( -160 \lambda - \frac{13}{3} g_1^2 +  \frac{28}{3}  g_2^2\right) \coef{6}{H^4\Box}{} \coef{6}{H^4D^2}{} \nn
& -31 g_2^2 \ \coef{6}{H^4\Box}{} \coef{6}{W^2H^2}{} - 7 g_1^2 \ \coef{6}{H^4\Box}{} \coef{6}{B^2H^2}{} - \frac{19}{2} g_1 g_2 \ \coef{6}{H^4\Box}{} \coef{6}{WBH^2}{} 
	\nn &
+5 g_2^2 \ \coef{6}{H^4D^2}{} \coef{6}{W^2H^2}{} 
 -\frac{3}{2} g_1^2 \ \coef{6}{H^4D^2}{} \coef{6}{B^2H^2}{} -\frac{7}{12} g_1 g_2 \ \coef{6}{H^4D^2}{} \coef{6}{WBH^2}{} 
	\nn &
-50 g_2^2 \left(\coef{6}{W^2H^2}{}\right)^2 - 14 g_1^2 \left(\coef{6}{B^2H^2}{}\right)^2 
+ \left( 32 \lambda -\frac{11}{2} g_1^2 -\frac{25}{6} g_2^2\right) \left( \coef{6}{WBH^2}{}\right)^2 
	\nn &
-11 g_1 g_2 \ \coef{6}{W^2H^2}{} \coef{6}{WBH^2}{} -17 g_1 g_2 \ \coef{6}{B^2H^2}{} \coef{6}{WBH^2}{} \nn
& -20 \ \coef{8}{H^8}{} + \left( -22 \lambda + \frac{1}{12} g_1^2 -\frac{17}{12} g_2^2 \right) \coef{8}{H^6D^2}{(1)} + \left( \frac{1}{6} g_1^2+ 3 g_2^2 \right) \coef{8}{H^6D^2}{(2)} \nn
& -9 g_2^2 \ \coef{8}{W^2H^4}{(1)} - 3 g_2^2 \ \coef{8}{W^2H^4}{(3)} -3 g_1^2 \ \coef{8}{B^2H^4}{(1)} -3 g_1 g_2 \ \coef{8}{WBH^4}{(1)} 
\biggr\} \,.
\end{align}
%
%
\subsubsection{$H^4D^2$}

The RGEs for the $H^4D^2$ couplings, which enter in the scalar metric, are
\begin{align}
	\dcoef{6}{H^4\Box}{} = m_H^2 & \biggl\{-24 \left(\coef{6}{H^4\Box}{}\right)^2 + \frac{3}{4} \left(\coef{6}{H^4D^2}{} \right)^2 + 8\ \ \coef{6}{H^4\Box}{} \coef{6}{H^4D^2}{} -64 \left(\coef{6}{G^2H^2}{}\right)^2 \nn
&  -24 \left(\coef{6}{W^2H^2}{}\right)^2- 8 \left(\coef{6}{B^2H^2}{}\right)^2 + 4  \left(\coef{6}{WBH^2}{}\right)^2 - 3\ \ \coef{8}{H^6D^2}{(1)} +2\ \ \coef{8}{H^6D^2}{(2)}
\biggr\} \,,
\end{align}
\begin{align}
\dcoef{6}{H^4 D^2}{} = & m_H^2 \biggl\{ 6 \left({\coef{6}{H^4D^2}{}}\right)^2- 8\ \ \coef{6}{H^4\Box}{} \coef{6}{H^4D^2}{}  -16 \left( \coef{6}{WBH^2}{} \right)^2 - 8\ \ \coef{8}{H^6D^2}{(2)}
\biggr\} \,.
\end{align}
%
%

\subsubsection{$X^2H^2$}

The RGEs for the $X^2H^2$ couplings, which enter in the gauge metric, are
\begin{align}
\dcoef{6}{G^2H^2}{}  =& m_H^2 \biggl\{ -14\ \ \coef{6}{H^4\Box}{} \coef{6}{G^2H^2}{} + 4\ \ \coef{6}{H^4D^2}{} \coef{6}{G^2H^2}{} -12 \left(\coef{6}{G^2H^2}{} \right)^2   - 6\ \ \coef{8}{G^2H^4}{(1)} 
\biggr\} \,,
\end{align}
\begin{align}
\dcoef{6}{W^2H^2}{}  =& m_H^2 \biggl\{ -14\ \ \coef{6}{H^4\Box}{} \coef{6}{W^2H^2}{} + 4\ \ \coef{6}{H^4D^2}{} \coef{6}{W^2H^2}{} -12 \left(\coef{6}{W^2H^2}{} \right)^2 - \left( \coef{6}{WBH^2}{} \right)^2 \nn  
& \qquad - 6\ \ \coef{8}{W^2H^4}{(1)} -2\ \  \coef{8}{W^2H^4}{(3)} 
\biggr\} \,,
\end{align}
\begin{align}
\dcoef{6}{B^2H^2}{} =& m_H^2 \biggl\{ -14\ \ \coef{6}{H^4\Box}{} \coef{6}{B^2H^2}{} + 4\ \ \coef{6}{H^4D^2}{} \coef{6}{B^2H^2}{} -12  \left(\coef{6}{B^2H^2}{} \right)^2 -3 \left(\coef{6}{WBH^2}{} \right)^2  
\nn & \qquad - 6\ \ \coef{8}{B^2H^4}{(1)} 
\biggr\} \,,
\end{align}
\begin{align}
\dcoef{6}{WBH^2}{}  =& m_H^2 \biggl\{ -6\ \ \coef{6}{H^4\Box}{} \coef{6}{WBH^2}{} + 2 \ \ \coef{6}{H^4D^2}{} \coef{6}{WBH^2}{} -8\ \ \coef{6}{W^2H^2}{} \coef{6}{WBH^2}{}  \nn
& \qquad -8\ \  \coef{6}{B^2H^2}{} \coef{6}{WBH^2}{} - 4\ \ \coef{8}{WBH^4}{(1)} 
\biggr\} \,. 
\end{align}
%
%
\subsection{Dimension 8}

The dimension-eight RGEs for the dimension-eight coefficients in the SMEFT Lagrangian are listed below. The contributions are all of order $1/M^4$ in the SMEFT power counting.

\subsubsection{$X^4$}

There are 26 even-parity $X^4$ operators at dimension eight~\cite{Murphy:2020rsh}. The non-zero RGEs involving the coefficients in Eq.~\eqref{BB.2} are:
\begin{align}
\dcoef{8}{G^4}{(1)} &= -2 \left(\coef{6}{G^2H^2}{}\right)^2 \,,
\end{align}
\begin{align}
\dcoef{8}{W^4}{(1)} &= -2  \left(\coef{6}{W^2H^2}{} \right)^2 \,,
\end{align}
\begin{align}
\dcoef{8}{B^4}{(1)} &= -2 \left(\coef{6}{B^2H^2}{} \right)^2 \,,
\end{align}
\begin{align}
\dcoef{8}{G^2W^2}{(1)} &= -4\ \  \coef{6}{G^2H^2}{} \coef{6}{W^2H^2}{}  \,,
\end{align}
\begin{align}
\dcoef{8}{G^2B^2}{(1)}  &= -4\ \ \coef{6}{G^2H^2}{} \coef{6}{B^2H^2}{}  \,,
\end{align}
\begin{align}
\dcoef{8}{W^2B^2}{(1)}  &= -4\ \ \coef{6}{W^2H^2}{} \coef{6}{B^2H^2}{}  \,,
\end{align}
\begin{align}
\dcoef{8}{W^2B^2}{(3)} &= -2 \left(\coef{6}{WBH^2}{} \right)^2 \,.
\end{align}
All other anomalous dimensions do not depend on the couplings in the Lagrangian in Eq.~\eqref{eq:Lagr}.

\subsubsection{$H^8$}

The RGE for the $H^8$ coupling is
\begin{align}\label{C.21}
\dcoef{8}{H^8}{}  = & -126 \left(\coef{6}{H^6}{}\right)^2 + \left( 960 \lambda - 20 g_2^2 \right) \coef{6}{H^6}{} \coef{6}{H^4\Box}{} + \left( -228 \lambda - 9 g_1^2+ 9 g_2^2\right) \coef{6}{H^6}{} \coef{6}{H^4D^2}{} \nn &
-54 g_2^2 \ \coef{6}{H^6}{} \coef{6}{W^2H^2}{} -18 g_1^2 \ \coef{6}{H^6}{} \coef{6}{B^2H^2}{} -18 g_1 g_2 \ \coef{6}{H^6}{} \coef{6}{WBH^2}{} \nn
&+ \left( -1568 \lambda^2 -\frac{40}{3}  \lambda  g_1^2+ 40 \lambda g_2^2  \right) \left(\coef{6}{H^4\Box}{}\right)^2  \nn
& + \left(-74 \lambda^2 -\frac{37}{12} \lambda g_1^2 + \frac{5}{4} \lambda g_2^2 -\frac{3}{8} g_1^4 -\frac{3}{4} g_1^2 g_2^2 - \frac{3}{8} g_2^4\right) \left(\coef{6}{H^4D^2}{}\right)^2 \nn
& + \left( 736 \lambda^2 + \frac{52}{3} \lambda g_1^2 - \frac{112}{3} \lambda g_2^2\right) \coef{6}{H^4\Box}{} \coef{6}{H^4D^2}{} + 124 \lambda g_2^2 \ \coef{6}{H^4\Box}{} \coef{6}{W^2H^2}{}
\nn &
+ 28 \lambda g_1^2 \ \coef{6}{H^4\Box}{}\coef{6}{B^2H^2}{} 
 + 38 \lambda g_1 g_2 \ \coef{6}{H^4\Box}{} \coef{6}{WBH^2}{}  
\nn &
+ \left(-20 \lambda g_2^2 -6 g_1^2 g_2^2 - 6 g_2^4 \right) \coef{6}{H^4D^2}{} \coef{6}{W^2H^2}{} \nn
& + \left(6 \lambda g_1^2 -6 g_1^4 - 6 g_1^2 g_2^2 \right) \coef{6}{H^4D^2}{} \coef{6}{B^2H^2}{} 
\nn &
+ \left( \frac{7}{3} \lambda g_1 g_2 - 6 g_1^3 g_2 - 6 g_1 g_2^3\right) \coef{6}{H^4D^2}{} \coef{6}{WBH^2}{} \nn
& - 512 \lambda^2 \left(\coef{6}{G^2H^2}{} \right)^2 + \left( -192 \lambda^2 + 200 \lambda g_2^2 -12 g_1^2 g_2^2 -54 g_2^4 \right) \left(\coef{6}{W^2H^2}{} \right)^2 \nn
& + \left(-64 \lambda^2 + 56 \lambda g_1^2  -18 g_1^4 -12 g_1^2 g_2 \right) \left(\coef{6}{B^2H^2}{}\right)^2  \nn
& + \left(-32 \lambda^2 + 22 \lambda g_1^2 
+ \frac{50}{3} \lambda g_2^2 -3 g_1^4 -12 g_1^2 g_2^2 -3 g_2^4 \right) \left(\coef{6}{WBH^2}{}\right)^2 
\nn &
-12 g_1^2 g_2^2 \ \coef{6}{W^2H^2}{} \coef{6}{B^2H^2}{} 
 + \left( 44 \lambda g_1 g_2 -12 g_1^3 g_2 -24 g_1 g_2^3 \right) \coef{6}{W^2H^2}{} \coef{6}{WBH^2}{}
\nn &
+ \left( 68 \lambda g_1 g_2 -24 g_1^3 g_2 -12 g_1 g_2^3 \right) \coef{6}{B^2H^2}{} \coef{6}{WBH^2}{} 
 + \left( 192 \lambda - 6 g_1^2 -18 g_2^2\right) \coef{8}{H^8}{} 
\nn &
+ \left( 40 \lambda^2 -\frac{1}{3} \lambda g_1^2  + \frac{17}{3} \lambda g_2^2 - \frac{3}{4} g_1^4
-\frac{3}{2} g_1^2 g_2^2 -\frac{9}{4} g_2^4 \right) \coef{8}{H^6D^2}{(1)} \nn
& + \left( 48 \lambda^2 -\frac{2}{3} \lambda g_1^2 -12 \lambda g_2^2 - \frac{3}{4} g_1^4
-\frac{3}{2} g_1^2 g_2^2 + \frac{3}{4} g_2^4 \right) \coef{8}{H^6D^2}{(2)} \nn
& + \left( 36 \lambda g_2^2 - 3 g_1^2 g_2^2 - 9 g_2^4 \right) \coef{8}{W^2H^4}{(1)} + \left( 12\lambda g_2^2  - 3 g_1^2 g_2^2  - 3 g_2^4  \right) \coef{8}{W^2H^4}{(3)}  \nn
& + \left( 12 \lambda g_1^2  - 3 g_1^4 - 3 g_1^2 g_2^2  \right) \coef{8}{B^2H^4}{(1)} + \left( 12 \lambda g_1 g_2  - 3 g_1^3 g_2  - 3 g_1 g_2^3  \right) \coef{8}{WBH^4}{(1)}   \,.
\end{align}

\subsubsection{$H^6D^2$}

The RGEs for the $H^6D^2$ couplings, which enter in the scalar metric, are
\begin{align}
\dcoef{8}{H^6D^2}{(1)} =& -96\ \ \coef{6}{H^6}{} \coef{6}{H^4\Box}{} -12\ \ \coef{6}{H^6}{} \coef{6}{H^4D^2}{} + \left( 352 \lambda + 20 g_1^2+ \frac{20}{3} g_2^2 \right) \left(\coef{6}{H^4\Box}{}\right)^2  \nn
& + \left(-23 \lambda + \frac{1}{8} g_1^2+ \frac{161}{24} g_2^2 \right) \left(\coef{6}{H^4D^2}{}\right)^2 
+ \left(-64 \lambda - 2 g_1^2+  12 g_2^2 \right) \coef{6}{H^4\Box}{} \coef{6}{H^4D^2}{} \nn
& -22 g_2^2 \ \coef{6}{H^4\Box}{} \coef{6}{W^2H^2}{} + 6 g_1^2 \ \coef{6}{H^4\Box}{} \coef{6}{B^2H^2}{} - \frac{32}{3} g_1 g_2 \ \coef{6}{H^4\Box}{} \coef{6}{WBH^2}{} 
\nn &
+ 8 g_2^2 \ \coef{6}{H^4D^2}{} \coef{6}{W^2H^2}{}  
 + 6 g_1^2 \ \coef{6}{H^4D^2}{} \coef{6}{B^2H^2}{} + \frac{43}{3} g_1 g_2 \ \coef{6}{H^4D^2}{} \coef{6}{WBH^2}{} 
\nn &
+ 512 \lambda  \left(\coef{6}{G^2H^2}{}\right)^2 + \left( 192 \lambda + 4 g_2^2 \right) \left(\coef{6}{W^2H^2}{} \right)^2 
 + \left( 64 \lambda+ 12 g_1^2 \right) \left(\coef{6}{B^2H^2}{} \right)^2 
\nn &
+ \left( -3 g_1^2 - 3 g_2^2\right) \left(\coef{6}{WBH^2}{}\right)^2 + \frac{80}{3} g_1 g_2 \ \coef{6}{W^2H^2}{} \coef{6}{WBH^2}{} 
 + \frac{8}{3} g_1 g_2 \ \coef{6}{B^2H^2}{} \coef{6}{WBH^2}{} 
\nn &
+ \left( 68 \lambda + \frac{1}{2} g_1^2 -\frac{31}{6} g_2^2 \right) \coef{8}{H^6D^2}{(1)}  + \left( -8 \lambda + 7 g_1^2 + \frac{17}{3} g_2^2 \right) \coef{8}{H^6D^2}{(2)} \,,
\end{align}
\begin{align}
\dcoef{8}{H^6D^2}{(2)} =&  -18\ \ \coef{6}{H^6}{} \coef{6}{H^4D^2}{} + \frac{40}{3} g_1^2 \left(\coef{6}{H^4\Box}{}\right)^2 + \left(-26 \lambda - \frac{35}{12} g_1^2 +\frac{11}{3} g_2^2 \right) \left(\coef{6}{H^4D^2}{} \right)^2 \nn
&  + \left( 64 \lambda + \frac{20}{3} g_1^2+ \frac{20}{3} g_2^2 \right) \coef{6}{H^4\Box}{} \coef{6}{H^4D^2}{}
+ 20 g_1^2 \ \coef{6}{H^4\Box}{} \coef{6}{B^2H^2}{} 
\nn &
+ \frac{35}{3} g_1 g_2 \ \coef{6}{H^4\Box}{} \coef{6}{WBH^2}{} 
    + 12 g_2^2 \ \coef{6}{H^4D^2}{} \coef{6}{W^2H^2}{} -3 g_1^2 \ \coef{6}{H^4D^2}{} \coef{6}{B^2H^2}{} 
\nn &
- \frac{13}{6} g_1 g_2 \ \coef{6}{H^4D^2}{}\coef{6}{WBH^2}{} 
  + 16 g_1^2 \left(\coef{6}{B^2H^2}{}\right)^2
\nn &
 + \left( 32 \lambda -10 g_1^2 + \frac{10}{3} g_2^2 \right) \left(\coef{6}{WBH^2}{} \right)^2  + \frac{94}{3} g_1 g_2 \ \coef{6}{W^2H^2}{} \coef{6}{WBH^2}{} \nn
&  -\frac{14}{3} g_1 g_2 \ \coef{6}{B^2H^2}{} \coef{6}{WBH^2}{}  + \frac{10}{3} g_1^2 \ \coef{8}{H^6D^2}{(1)}
+ \left( 56 \lambda -\frac{7}{3} g_1^2+ \frac{10}{3} g_2^2 \right) \coef{8}{H^6D^2}{(2)} \,.
\end{align}

\subsubsection{$H^4D^4$}

The $H^4D^4$ operators are not included in the initial Lagrangian in Eq.~\eqref{eq:Lagr}, but they are generated in the counterterm structure. The RGEs are
\begin{align}
\dcoef{8}{H^4 D^4}{(1)}  =& -\frac{16}{3}\ \left(\coef{6}{H^4\Box}{} \right)^2 - \frac{11}{3} \left(\coef{6}{H^4D^2}{}\right)^2 + \frac{32}{3} \ \coef{6}{H^4\Box}{} \coef{6}{H^4D^2}{} -16 \left(\coef{6}{WBH^2}{}\right)^2 \,,
\end{align}
\begin{align}
\dcoef{8}{H^4 D^4}{(2)}  =& -\frac{16}{3} \left(\coef{6}{H^4\Box}{} \right)^2 - \frac{5}{3} \left( \coef{6}{H^4D^2}{}\right)^2 - \frac{16}{3} \ \coef{6}{H^4\Box}{} \coef{6}{H^4D^2}{} \,,
\end{align}
\begin{align}
\dcoef{8}{H^4 D^4}{(3)}  =& -\frac{40}{3} \left(\coef{6}{H^4\Box}{} \right)^2 +  \frac{7}{3} \left( \coef{6}{H^4D^2}{} \right)^2 - \frac{16}{3} \ \coef{6}{H^4\Box}{} \coef{6}{H^4D^2}{} - 128 \left(\coef{6}{G^2H^2}{} \right)^2  \nn
			  & -48 \left(\coef{6}{W^2H^2}{}\right)^2 
  - 16 \left(\coef{6}{B^2H^2}{}\right)^2 + 8 \left(\coef{6}{WBH^2}{}\right)^2 \,.
\end{align}
%

\subsubsection{$X^2H^4$}

The RGEs for the $X^2H^4$ coefficients, which enter in the gauge metric, are
\begin{align}
\dcoef{8}{G^2H^4}{(1)}  =& -24\ \ \coef{6}{H^6}{} \coef{6}{G^2H^2}{} + \left( 88 \lambda - \frac{20}{3} g_2^2 \right) \coef{6}{H^4\Box}{}\coef{6}{G^2H^2}{}
\nn &
+ \left( -24 \lambda - 3 g_1^2 + 3 g_2^2 \right) \coef{6}{H^4D^2}{} \coef{6}{G^2H^2}{} 
+ 48 \lambda \left(\coef{6}{G^2H^2}{} \right)^2 - 18 g_2^2 \ \coef{6}{G^2H^2}{} \coef{6}{W^2H^2}{} 
\nn &
- 6 g_1^2 \ \coef{6}{G^2H^2}{} \coef{6}{B^2H^2}{} - 6 g_1 g_2 \ \coef{6}{G^2H^2}{} \coef{6}{WBH^2}{} \nn
& + \left( 48 \lambda -3 g_1^2 -9 g_2^2 -22 g_3^2 \right)\coef{8}{G^2H^4}{(1)} \,,
\end{align}
\begin{align}
\dcoef{8}{W^2H^4}{(1)}  =& -24\ \ \coef{6}{H^6}{} \coef{6}{W^2H^2}{} -\frac{1}{8} g_2^2 \left({\coef{6}{H^4D^2}{}}\right)^2 + \left( 88 \lambda -2 g_2^2 \right) \coef{6}{H^4\Box}{}\coef{6}{W^2H^2}{} 
\nn &
+ \frac{5}{6} g_1 g_2 \ \coef{6}{H^4\Box}{} \coef{6}{WBH^2}{} 
 + \left( -24 \lambda - 3 g_1^2 + \frac{3}{2} g_2^2 \right) \coef{6}{H^4D^2}{} \coef{6}{W^2H^2}{} 
\nn &
 - \frac{5}{12} g_1 g_2 \ \coef{6}{H^4D^2}{} \coef{6}{WBH^2}{} + \left( 48 \lambda - \frac{38}{3} g_2^2\right) \left(\coef{6}{W^2H^2}{} \right)^2 \nn
& + \left( 4 \lambda + g_1^2 -\frac{13}{3} g_2^2\right) \left(\coef{6}{WBH^2}{}\right)^2  - 6 g_1^2 \ \coef{6}{W^2H^2}{} \coef{6}{B^2H^2}{} 
\nn &
- \frac{13}{3} g_1 g_2 \ \coef{6}{W^2H^2}{} \coef{6}{WBH^2}{} 
 - \frac{1}{3} g_1 g_2 \ \coef{6}{B^2H^2}{} \coef{6}{WBH^2}{}
\nn &
+ \left( 48 \lambda -3 g_1^2 -\frac{58}{3} g_2^2 \right) \coef{8}{W^2H^4}{(1)} + \left( 8 \lambda - 8 g_2^2 \right) \coef{8}{W^2H^4}{(3)}+ g_1 g_2 \ \coef{8}{WBH^4}{(1)}  \,,
\end{align}
\begin{align}
\dcoef{8}{W^2H^4}{(3)} =& \frac{19}{6} g_1 g_2 \ \coef{6}{H^4\Box}{} \coef{6}{WBH^2}{} + \frac{5}{2} g_2^2 \ \coef{6}{H^4D^2}{} \coef{6}{W^2H^2}{} - \frac{1}{12} g_1 g_2 \ \coef{6}{H^4D^2}{} \coef{6}{WBH^2}{} \nn
& + \left( 8 \lambda + 6 g_2^2 \right) \left( \coef{6}{WBH^2}{} \right)^2 
+ \frac{19}{3} g_1 g_2 \ \coef{6}{W^2H^2}{} \coef{6}{WBH^2}{} + \frac{1}{3} g_1 g_2 \ \coef{6}{B^2H^2}{} \coef{6}{WBH^2}{} \nn
& + \left( 24  \lambda -3 g_1^2+ \frac{14}{3} g_2^2 \right) \coef{8}{W^2H^4}{(3)} + g_1 g_2  \ \coef{8}{WBH^4}{(1)}  \,,
\end{align}
\begin{align}
\dcoef{8}{B^2H^4}{(1)}  =&  -24\ \ \coef{6}{H^6}{} \coef{6}{B^2H^2}{} + \frac{1}{8} g_1^2 \left(\coef{6}{H^4D^2}{} \right)^2 + \left( 88 \lambda + \frac{14}{3} g_1^2 - \frac{20}{3} g_2^2 \right) \coef{6}{H^4\Box}{} \coef{6}{B^2H^2}{} \nn
& + \frac{17}{3} g_1 g_2 \ \coef{6}{H^4\Box}{} \coef{6}{WBH^2}{}
+ \left( -24 \lambda - \frac{1}{3} g_1^2  + 3 g_2^2 \right) \coef{6}{H^4D^2}{} \coef{6}{B^2H^2}{}  
\nn &
+ 3 g_1 g_2 \ \coef{6}{H^4D^2}{} \coef{6}{WBH^2}{} 
 + \left( 48 \lambda - \frac{2}{3} g_1^2\right) \left(\coef{6}{B^2H^2}{} \right)^2
-18 g_2^2 \ \coef{6}{W^2H^2}{} \coef{6}{B^2H^2}{} 
\nn &
+ \left( 20 \lambda + 5 g_1^2 +  g_2^2 \right) \left( \coef{6}{WBH^2}{} \right)^2 
- \frac{2}{3} g_1 g_2 \ \coef{6}{W^2H^2}{} \coef{6}{WBH^2}{}
\nn &
+ \frac{16}{3} \ \coef{6}{B^2H^2}{} \coef{6}{WBH^2}{} +
\left( 48 \lambda + \frac{4}{3} g_1^2 - 9 g_2^2 \right) \coef{8}{B^2H^4}{(1)} + 4 g_1 g_2  \ \coef{8}{WBH^4}{(1)} \,,
\end{align}
\begin{align}
\dcoef{8}{WBH^4}{(1)} =&  -12\ \ \coef{6}{H^6}{} \coef{6}{WBH^2}{} + \frac{14}{3} g_1 g_2\ \coef{6}{H^4\Box}{} \coef{6}{W^2H^2}{} + \frac{14}{3} g_1 g_2\ \coef{6}{H^4\Box}{} \coef{6}{B^2H^2}{} \nn
&  + \left( 56 \lambda + 4 g_1^2 - g_2^2\right) \coef{6}{H^4\Box}{} \coef{6}{WBH^2}{}
+  g_1 g_2\ \coef{6}{H^4D^2}{} \coef{6}{W^2H^2}{} 
\nn &
+ \frac{8}{3} g_1 g_2\ \coef{6}{H^4D^2}{} \coef{6}{B^2H^2}{} 
  + \left(-16 \lambda - \frac{7}{2} g_1^2 + 6 g_2^2 \right) \coef{6}{H^4D^2}{} \coef{6}{WBH^2}{} 
\nn &
+ \frac{4}{3} g_1 g_2 \left(\coef{6}{W^2H^2}{} \right)^2 
+ \frac{4}{3} g_1 g_2 \left(\coef{6}{B^2H^2}{}\right)^2 
 + \frac{8}{3} g_1 g_2 \left(\coef{6}{WBH^2}{} \right)^2 
\nn &
+ 8 g_1 g_2 \ \coef{6}{W^2H^2}{} \coef{6}{B^2H^2}{} + \left(48 \lambda + 4  g_1^2 + \frac{28}{3} g_2^2\right) \coef{6}{W^2H^2}{} \coef{6}{WBH^2}{}  \nn
& + \left( 48 \lambda - 2 g_1^2 + \frac{22}{3} g_2^2\right) \coef{6}{B^2H^2}{} \coef{6}{WBH^2}{} + 4 g_1 g_2 \ \coef{8}{W^2H^4}{(1)} + 8 g_1 g_2 \ \coef{8}{W^2H^4}{(3)}
\nn &
+ 4 g_1 g_2 \ \coef{8}{B^2H^4}{(1)}  + \left( 40 \lambda -\frac{5}{6} g_1^2 - \frac{37}{6} g_2^2 \right)  \coef{8}{WBH^4}{(1)}  \,.
\end{align}
%
%

\subsubsection{$X^3H^2$}

The $X^3H^2$ operators are not included in the initial Lagrangian in Eq.~\eqref{eq:Lagr}, but they are generated in the counterterm structure. The RGEs are
\begin{align}
\dcoef{8}{G^3H^2}{(1)} &=  16 g_3 \left(\coef{6}{G^2H^2}{}\right)^2 \,,
\end{align}
\begin{align}
\dcoef{8}{W^3H^2}{(1)} &=  16 g_2 \left(\coef{6}{W^2H^2}{}\right)^2 \,,
\end{align}
\begin{align}
\dcoef{8}{W^2BH^2}{(1)} &=  16 g_2 \coef{6}{W^2H^2}{} \coef{6}{WBH^2}{} \,.
\end{align}
%
%

\subsubsection{$X^2H^2D^2$}

The $X^2H^2D^2$ operators are not included in the initial Lagrangian in Eq.~\eqref{eq:Lagr}, but they are generated in the counterterm structure. The RGEs are
\begin{align}
\dcoef{8}{G^2H^2D^2}{(1)}  &=  \frac{32}{3} \left(\coef{6}{G^2H^2}{}\right)^2 \,,
\end{align}
\begin{align}
\dcoef{8}{G^2H^2D^2}{(2)}  &=  -12\ \ \coef{6}{H^4\Box}{} \coef{6}{G^2H^2}{} - \frac{8}{3} \left(\coef{6}{G^2H^2}{} \right)^2 \,,
\end{align}
\begin{align}
\dcoef{8}{W^2H^2D^2}{(1)}  &=  \frac{32}{3} \left(\coef{6}{W^2H^2}{} \right)^2 + \frac{8}{3} \left(\coef{6}{WBH^2}{} \right)^2 \,,
\end{align}
\begin{align}
\dcoef{8}{W^2H^2D^2}{(2)}  &=  -12 \ \ \coef{6}{H^4\Box}{} \coef{6}{W^2H^2}{} - \frac{8}{3} \left(\coef{6}{W^2H^2}{} \right)^2 - \frac{2}{3} \left( \coef{6}{WBH^2}{} \right)^2 \,,
\end{align}
\begin{align}
\dcoef{8}{W^2H^2D^2}{(4)}  &=  0 \,,
\end{align}

\begin{align}
\dcoef{8}{B^2H^2D^2}{(1)}  &=  \frac{32}{3} \left(\coef{6}{B^2H^2}{} \right)^2 + \frac{8}{3} \left(\coef{6}{WBH^2}{} \right)^2 \,,
\end{align}
\begin{align}
\dcoef{8}{B^2H^2D^2}{(2)} &= -12\ \ \coef{6}{H^4\Box}{} \coef{6}{B^2H^2}{} - \frac{8}{3} \left(\coef{6}{B^2H^2}{} \right)^2 -  2  \left( \coef{6}{WBH^2}{} \right)^2 \,,
\end{align}
\begin{align}
\dcoef{8}{WBH^2D^2}{(1)} =& 4\ \ \coef{6}{H^4\Box}{} \coef{6}{WBH^2}{}  -4\ \ \coef{6}{H^4D^2}{} \coef{6}{WBH^2}{} - \frac{8}{3}\ \coef{6}{W^2H^2}{} \coef{6}{WBH^2}{} 
\nn &
- \frac{8}{3}\ \coef{6}{B^2H^2}{} \coef{6}{WBH^2}{} \,,
\end{align}
\begin{align}
\dcoef{8}{WBH^2D^2}{(3)} &= 0 \,,
\end{align}
\begin{align}
\dcoef{8}{WBH^2D^2}{(4)} &= \frac{16}{3} \ \coef{6}{W^2H^2}{} \coef{6}{WBH^2}{} +  \frac{16}{3} \ \coef{6}{B^2H^2}{} \coef{6}{WBH^2}{} \,.
\end{align}
%
%

\subsubsection{$XH^4D^2$}

The $XH^4D^2$ operators are not included in the initial Lagrangian in Eq.~\eqref{eq:Lagr}, but they are generated in the counterterm structure. The RGEs are
\begin{align}
\dcoef{8}{BH^4D^2}{(1)} =&  -g_1  \left( \coef{6}{H^4D^2}{}\right)^2 + \frac{40}{3} g_1\  \coef{6}{H^4\Box}{} \coef{6}{B^2H^2}{} - 20 g_2\ \coef{6}{H^4\Box}{} \coef{6}{WBH^2}{}  
\nn &
-  \frac{32}{3} g_1 \ \coef{6}{H^4D^2}{} \coef{6}{B^2H^2}{} 
  -12 g_2 \ \coef{6}{H^4D^2}{} \coef{6}{WBH^2}{} + \frac{80}{3} g_1 \left(\coef{6}{B^2H^2}{}\right)^2 
\nn &
- 28 g_1 \left(\coef{6}{WBH^2}{} \right)^2 + 56 g_2 \ \coef{6}{W^2H^2}{} \coef{6}{WBH^2}{} 
 -40 g_2 \ \coef{6}{B^2H^2}{} \coef{6}{WBH^2}{} \,,
\end{align}
\begin{align}
\dcoef{8}{WH^4D^2}{(1)} =& g_2 \left(\coef{6}{H^4D^2}{}\right)^2 + \frac{40}{3} g_2 \ \coef{6}{H^4\Box}{} \coef{6}{W^2H^2}{} - \frac{20}{3} g_1 \ \coef{6}{H^4\Box}{} \coef{6}{WBH^2}{} 
\nn &
+ \frac{28}{3} g_1 \ \coef{6}{H^4D^2}{} \coef{6}{WBH^2}{} 
 + \frac{80}{3} g_2 \left(\coef{6}{W^2H^2}{}\right)^2 - 4 g_2 \left(\coef{6}{WBH^2}{}\right)^2 
\nn &
- \frac{40}{3} g_1 \ \coef{6}{W^2H^2}{} \coef{6}{WBH^2}{}  + \frac{56}{3} g_1 \ \coef{6}{B^2H^2}{} \coef{6}{WBH^2}{} \,,
\end{align}
\begin{align}
\dcoef{8}{WH^4D^2}{(3)} =& 0 \,.
\end{align}

\bibliographystyle{JHEP}
\bibliography{bibliographyManifold.bib}

\end{document}